\documentclass{article}

\usepackage[final,eandd]{neurips_2026}

\usepackage[utf8]{inputenc}
\usepackage[T1]{fontenc}
\usepackage{hyperref}
\usepackage{url}
\usepackage{booktabs}
\usepackage{amsfonts}
\usepackage{amsmath}
\usepackage{nicefrac}
\usepackage{microtype}
\usepackage{xcolor}
\usepackage{graphicx}
\usepackage{subcaption}
\usepackage{multirow}
\usepackage{float}
\newfloat{algorithm}{t}{loa}
\floatname{algorithm}{Algorithm}
\usepackage{enumitem}
\usepackage{colortbl}
\usepackage{array}
\usepackage{xspace}
\usepackage{pifont}
\usepackage{wrapfig}

\newcommand{\bench}{AV-Phys Bench\xspace}
\newcommand{\agent}{AV-Phys Agent\xspace}

\definecolor{tabgreen}{rgb}{0.20,0.55,0.27}
\definecolor{tabred}{rgb}{0.78,0.18,0.18}
\definecolor{taborange}{rgb}{0.92,0.55,0.10}
\definecolor{rowprop}{RGB}{230,245,240}
\definecolor{rowopen}{RGB}{242,248,252}
\newcommand{\cmark}{\textcolor{tabgreen}{\ding{51}}}
\newcommand{\xmark}{\textcolor{tabred}{\ding{55}}}

\newcommand{\vsa}{\texttt{V-SA}\xspace}
\newcommand{\asa}{\texttt{A-SA}\xspace}
\newcommand{\vpc}{\texttt{V-PC}\xspace}
\newcommand{\apc}{\texttt{A-PC}\xspace}
\newcommand{\avpc}{\texttt{AV-PC}\xspace}

\usepackage{fontawesome5}
\definecolor{linkrowdark}{HTML}{222222}
\newcommand{\plainlink}[2]{{\hypersetup{hidelinks}\href{#1}{#2}}}
\newcommand{\hflogo}{\raisebox{-0.2em}{\includegraphics[height=1.15em]{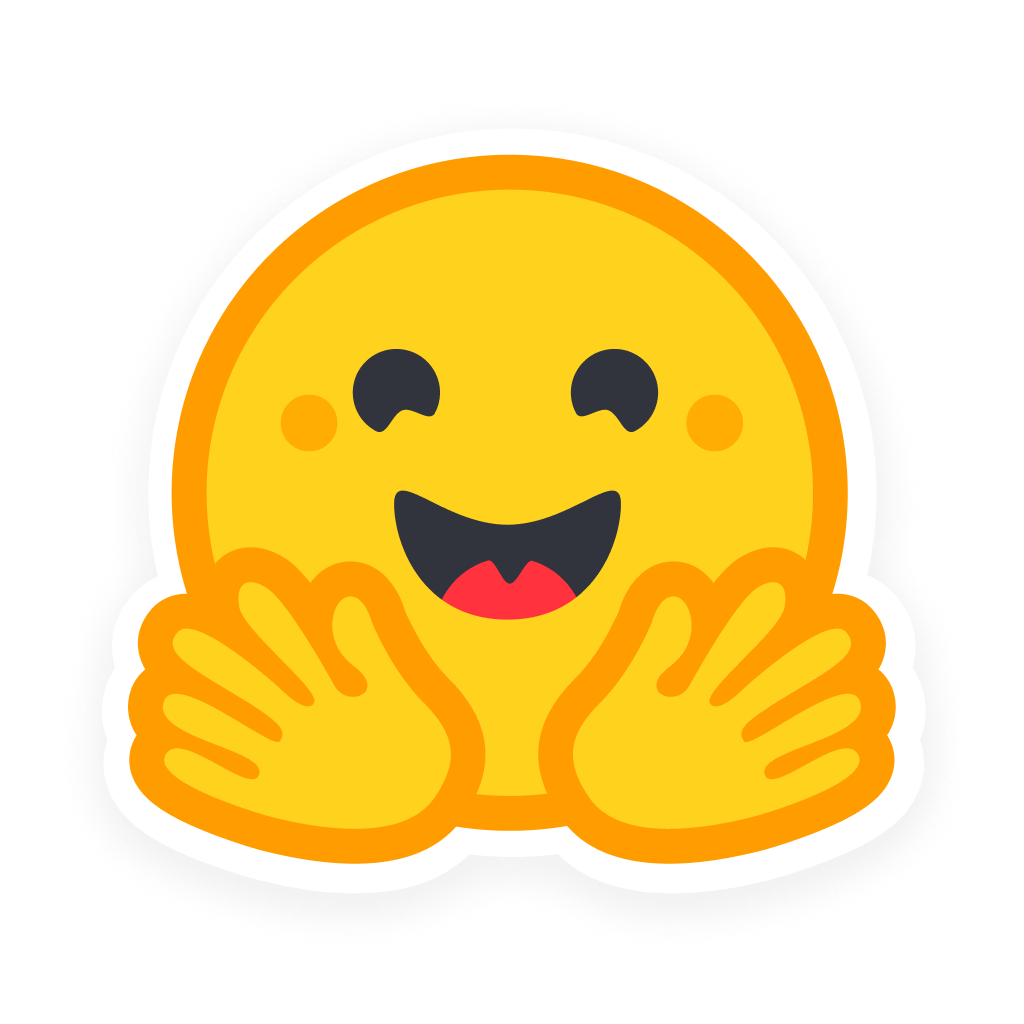}}}

\title{Do Joint Audio-Video Generation Models \\ Understand Physics?}

\author{%
  \makebox[\textwidth][s]{%
    Zijun Cui$^{1,*}$\hfil Xiulong Liu$^{2,*}$\hfil Hao Fang$^{2,*}$\hfil Mingwei Xu$^{2}$\hfil Jiageng Liu$^{3}$%
  } \\
  \textbf{Zexin Xu$^{1}$ \quad Weiguo Pian$^{1}$ \quad Shijian Deng$^{1}$ \quad
  Feiyu Du$^{1}$ \quad Chenming Ge$^{2}$ \quad Yapeng Tian$^{1,\dagger}$} \\[4pt]
  $^{1}$University of Texas at Dallas \enspace
  $^{2}$University of Washington \enspace
  $^{3}$University of California, Los Angeles \\[2pt]
  $^{*}$Equal contribution. \quad $^{\dagger}$Corresponding author. \\[5pt]
  \plainlink{https://huggingface.co/datasets/ZijunCui/AV-Phys-Bench}{\hflogo\,{\color{linkrowdark}Dataset}}\quad\quad \plainlink{https://github.com/ZijunCui02/AV-Phys}{{\color{linkrowdark}\faGithub\,Code}}\quad\quad \plainlink{https://zijuncui.com/AV-Phys/}{{\color{linkrowdark}\faGlobe\,Homepage}}
}

\begin{document}
\addtocontents{toc}{\protect\setcounter{tocdepth}{0}}

\setlength{\tabcolsep}{0pt}
\maketitle
\setlength{\tabcolsep}{6pt}
\vspace{-9.5mm}

\begin{figure}[H]
  \centering
  \includegraphics[width=0.98\linewidth]{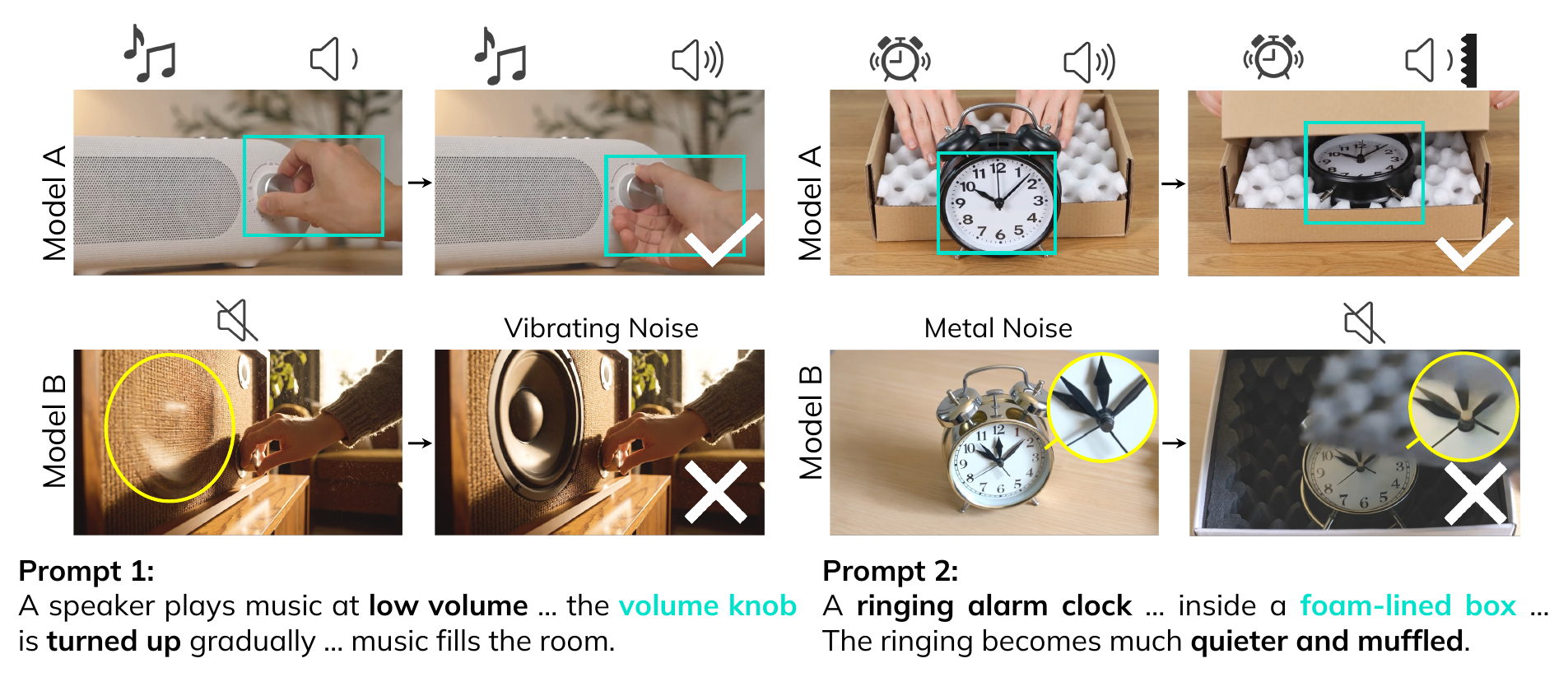}
  \vspace{-1mm}
  \caption{\textbf{If you turn up the volume knob in an AI-generated video, does the music get louder?} In the physical world, vision and sound are two observations of the same physical event. 
  For example, turning up a volume knob makes the music louder, while placing a ringing clock inside a foam-lined box muffles its sound.
  \textbf{Top:}  A SOTA joint audio-video generation model~\cite{seedance2026seedance} correctly captures these physical effects in both video and audio.
  \textbf{Bottom:} Another SOTA model~\cite{team2025kling} fails to preserve physical consistency: it hallucinates a previously absent speaker membrane and generates vibration noise, and renders a six-handed clock whose ringing disappears after the box is closed.}
  \label{fig:teaser}
\end{figure}

\begin{abstract}
Joint audio-video generation models are rapidly approaching professional production quality, raising a fundamental question: \textit{do these models truly understand audio-visual physics, or merely generate plausible sounds and frames that violate real-world physical consistency?}
To answer this question, in this paper, we introduce \bench, the first benchmark for evaluating physical commonsense in joint audio-video generation.
\bench systematically tests joint audio-video generation models across three scene categories that probe how physical commonsense holds as the scene evolves: (a) Steady State, (b) Event Transition, and (c) Environment Transition. Each scene category investigates three physics-grounded subcategories that reflect real-world scenes, and an additional Anti-AV-Physics subcategory where prompts deliberately violate audio-visual physics to probe whether models possess generative physics knowledge or merely encode physically consistent priors. Each generation is scored along five dimensions: semantic adherence and physical commonsense within each modality, and a cross-modal physical commonsense dimension that tests whether the visual and audio streams agree on the same physical event. Across three proprietary and four open-source models, Seedance 2.0 leads on physical commonsense (overall pass rate $0.660$), the gap to open-source models remains pronounced, proprietary models degrade by up to $61\%$ on event-driven and environment-driven transitions, and proprietary leaders collapse by $45$--$69\%$ on Anti-AV-Physics prompts. Beyond the human evaluation, we introduce \agent, a ReAct-style agentic evaluator that pairs a multimodal language model with deterministic acoustic measurement tools and ranks generators in close alignment with human ratings, enabling scalable physical-commonsense evaluation without further human annotation. \bench identifies cross-modal physical consistency and transition-driven scene dynamics as the open frontier for joint audio-video generation. We release the \bench and \agent for future facilitation of joint audio-video generation and evaluations.
\end{abstract}

\section{Introduction}
\label{sec:intro}

Recent advances in joint audio-video generation ~\cite{seedance2026seedance,team2025kling,sora2,veo31,hacohen2026ltx,low2025ovi} have enabled models to synthesize highly realistic videos together with synchronized sound. As these systems become increasingly perceptually convincing, an important challenge emerges: realism alone does not guarantee physical consistency. In the physical world, visual and acoustic signals are governed by the same underlying dynamics. Actions such as turning a volume knob or enclosing a ringing clock inside a foam-lined box simultaneously alter both what we see and what we hear. A reliable joint generation model should therefore preserve coherent physical relationships across modalities as scenes evolve over time. However, as illustrated in Figure~\ref{fig:teaser}, current state-of-the-art models can still produce physically inconsistent generations, including implausible visual artifacts and audio that contradicts the depicted scene dynamics. These failures suggest that generating perceptually realistic audio and video is fundamentally different from modeling the causal physical relationships that jointly govern both modalities. This distinction is particularly important for downstream applications such as world simulation~\cite{agarwal2025cosmos, bruce2024genie, brooks2024video}, embodied agents~\cite{chen2022soundspaces,gan2020look, liu2024caven, chen2025savvy}, and educational content, where maintaining physically consistent audio-visual behavior is essential.

Evaluating joint audio-video generation, therefore, requires a benchmark that goes beyond perceptual quality and semantic alignment to test whether audio, video, and their interaction remain physically consistent. Existing benchmarks have made important progress on related aspects. PhysicsIQ~\cite{motamed2026generative}, PhyGenBench~\cite{meng2024towards}, VideoPhy-2~\cite{bansal2025videophy}, and PhyWorldBench~\cite{gu2025phyworldbench} focus on physical realism in video, while PhyAVBench~\cite{xie2025phyavbench} examines whether generated audio responds appropriately to controlled changes in material, force, and environment. TAVGBench~\cite{mao2024tavgbench}, JavisBench~\cite{liu2025javisdit}, VABench~\cite{hua2025vabench}, and SAVGBench~\cite{shimada2026savgbench} instead evaluate semantic, temporal, and spatial alignment in joint audio-video generation. Across them, however, physics is typically assessed within a single isolated event, leaving open whether models can sustain physical commonsense as a scene evolves through an action or environmental change. More critically, it remains unclear whether models truly understand the audio-visual physics of the events they depict. To answer these questions, we introduce \bench, the first benchmark for audio-visual physical consistency in joint audio-video generation.

\bench is designed to evaluate whether joint audio-video models sustain physical commonsense as scenes evolve. To this end, it organizes evaluation into three scene categories corresponding to different types of scene dynamics: \emph{Steady State}, \emph{Event Transition}, and \emph{Environment Transition}. \emph{Steady State} covers scenes whose physical configuration remains unchanged over time, extending prior single-event physics evaluation to the audio-visual setting. For example, a spinning coin should produce both a stable visual rotation and a corresponding metallic ringing sound. \emph{Event Transition} tests scenes in which a discrete action should produce a clear audio-visual change. For example, when a hand turns a volume knob, the model should update both the visual state and the sound accordingly. \emph{Environment Transition} tests scenes in which changing the surroundings should alter the perceived event. For example, placing an alarm clock inside a foam-lined box should change both how the scene looks and how the ringing sounds. In addition, each scene category includes an \emph{Anti-AV-Physics} subcategory, where prompts deliberately violate audio-visual physics to test whether models capture underlying physical rules rather than simply generating physically plausible outputs. To assess model behavior in each scene category, \bench scores every output along five complementary dimensions: visual semantic adherence, visual physical commonsense, audio semantic adherence, audio physical commonsense, and cross-modal physical commonsense.

\begin{table}[h]
\vspace{-2.6mm}
\centering
\small
\setlength{\tabcolsep}{6pt}
\renewcommand{\arraystretch}{1.0}
\resizebox{\textwidth}{!}{%
\begin{tabular}{@{}lcccccc@{}}
\toprule
Benchmark & V Semantic & A Semantic & V Phys. & A Phys. & AV Phys. & Scene Evolutions \\
\midrule
VBench~\cite{huang2024vbench}            & \cmark          & \xmark          & \xmark          & \xmark          & \xmark          & \xmark \\
TAVGBench~\cite{mao2024tavgbench}        & \cmark          & \cmark          & \xmark          & \xmark          & \xmark          & \xmark \\
VABench~\cite{hua2025vabench}            & \cmark          & \cmark          & \cmark          & \cmark          & \xmark          & \xmark \\
JavisBench~\cite{liu2025javisdit}        & \cmark          & \cmark          & \xmark          & \xmark          & \xmark          & \xmark \\
PhysicsIQ~\cite{motamed2026generative}   & \xmark          & \xmark          & \cmark          & \xmark          & \xmark          & \xmark \\
PhyGenBench~\cite{meng2024towards}       & \cmark          & \xmark          & \cmark          & \xmark          & \xmark          & \cmark \\
VideoPhy-2~\cite{bansal2025videophy}     & \cmark          & \xmark          & \cmark          & \xmark          & \xmark          & \cmark \\
PhyWorldBench~\cite{gu2025phyworldbench} & \cmark          & \xmark          & \cmark          & \xmark          & \xmark          & \xmark \\
PhyAVBench~\cite{xie2025phyavbench}      & \cmark          & \cmark          & \xmark          & \cmark          & \xmark          & \xmark \\
\textbf{\bench (Ours)}                   & \textbf{\cmark} & \textbf{\cmark} & \textbf{\cmark} & \textbf{\cmark} & \textbf{\cmark} & \textbf{\cmark} \\
\bottomrule
\end{tabular}}
\vspace{2mm}
\caption{\textbf{Comparison of Benchmarks.} V/A Semantic and V/A Phys. mark whether the benchmark scores semantic adherence and physical commonsense within each modality. AV Phys. marks cross-modal physical consistency between the two streams. Scene Evolutions marks whether the prompt set deliberately covers event or environment transitions rather than only single isolated events. VABench's V/A Phys. are minor sub-axes within broader realism metrics; \bench is the only benchmark to cover all six axes and the only one to score AV Phys.}
\vspace{-5.6mm}
\label{tab:comparison}
\end{table}

Our human evaluation collects reliable annotations across all five dimensions, providing the gold-standard reference for \bench. For automation at scale, we introduce \agent, an MLLM-based pipeline built on Gemini~\cite{comanici2025gemini} that jointly scores each generation's visual and audio content along the same five dimensions, validated against the human ratings. \agent is a ReAct-style evaluator that augments multimodal LLM reasoning with deterministic audio tools. \agent resolves semantic rubric statements from direct perception, while grounding physics-sensitive statements in targeted tools. By feeding these measurements back to the model as structured evidence, \agent produces scalable rubric-based judgments that align more closely with human ratings than standard MLLM-as-judge evaluation.

To assess current joint audio-video generative systems on physical commonsense, we evaluate seven recent proprietary and open-source models on \bench. Using our rubric framework, we analyze performance across scene categories, modalities, and Anti-AV-Physics prompts. Our experiments reveal four main findings: current systems exhibit a pronounced semantics-to-physics gap, with performance degrading from semantic adherence to unimodal physics and further to cross-modal physical consistency; proprietary models substantially outperform open-source ones, with distinct modality-level error profiles; scene transitions emerge as the main difficulty frontier; and Anti-AV-Physics prompts expose sharp failures even in the strongest proprietary model. We further show that our \agent design aligns more closely with human judgments than other baselines.

To summarize, our contributions are: 1) We introduce \bench, a first-of-its-kind comprehensive benchmark for evaluating physical commonsense in joint audio-video generation models; 2) We build a rubric-based evaluation framework around physics-grounded prompts, with rubric criteria tailored to each prompt for evaluating physical commonsense across multiple axes; 3) We propose \agent, an evaluator that combines multimodal reasoning with deterministic audio-DSP tools for scalable rubric-based evaluation aligned with human judgments; 4) We conduct comprehensive experiments on \bench, which reveal the strengths and limitations of current joint audio-video generation models.

\section{Related Work}
\label{sec:related}

\textbf{Physics-aware evaluation.}
Recent joint audio-video generation models, including proprietary systems such as Seedance 2.0~\cite{seedance2026seedance}, Kling 3.0 Omni~\cite{team2025kling}, Veo 3.1~\cite{veo31}, and Sora 2~\cite{sora2}, as well as open-source models such as Ovi~\cite{low2025ovi}, JavisDiT++~\cite{liu2026javisdit++}, LTX-2.3~\cite{hacohen2026ltx}, and UniVerse-1~\cite{wang2025universe}, can generate video with synchronized audio. Their progress has motivated a growing line of physics-aware evaluation, though existing benchmarks remain limited in scope. On the visual side, benchmarks such as PhysicsIQ~\cite{motamed2026generative}, PhyGenBench~\cite{meng2024towards}, VideoPhy~\cite{bansal2024videophy}, VideoPhy-2~\cite{bansal2025videophy}, PhyWorldBench~\cite{gu2025phyworldbench}, and T2VPhysBench~\cite{guo2025t2vphysbench} evaluate physical realism or physical commonsense in generated video. On the audio side, PhyAVBench~\cite{xie2025phyavbench} tests whether generated audio responds appropriately to controlled physical changes, but does not consider visual physics or joint scene dynamics. More general video or audio-video benchmarks, such as VBench~\cite{huang2024vbench}, primarily evaluate perceptual quality rather than physical consistency. In contrast, our benchmark evaluates acoustic physics, visual physics, and cross-modal physical consistency jointly, including whether physical commonsense is preserved as scenes evolve through actions and environment changes.

\textbf{Audio-visual generation benchmarks.}
Existing audio-visual generation benchmarks evaluate how well models jointly produce video and audio, but they primarily focus on semantic, temporal, or spatial alignment rather than physical commonsense. TAVGBench~\cite{mao2024tavgbench}, JavisBench~\cite{liu2025javisdit}, SAVGBench~\cite{shimada2026savgbench}, Verse-Bench~\cite{wang2025universe}, VABench~\cite{hua2025vabench}, and T2AV-Compass~\cite{cao2025t2av} each target different aspects of audio-video correspondence, such as semantic relevance, synchronization, or spatial consistency. In contrast, our benchmark asks whether generated audio and video are not only aligned, but also physically consistent with the same depicted event and its evolution over time. Table~\ref{tab:comparison} compares \bench with prior representative benchmarks in the dimensions that we cover.

\textbf{Evaluation metrics.}
Existing metrics remain limited in their ability to evaluate physical commonsense in joint audio-video generation.
Distributional metrics such as FVD~\cite{unterthiner2018towards}, FAD~\cite{kilgour2019reference}, and CLIP-based scores~\cite{hessel2021clipscore} capture aggregate similarity to reference data, but they do not localize specific physical violations. Alignment metrics such as AV-Align~\cite{yariv2024diverse} measure cross-modal synchronization, but not whether the generated audio and video are physically correct. Human evaluation remains the gold standard and is widely used in physics-aware benchmarks~\cite{bansal2024videophy,motamed2026generative,xie2025phyavbench}, but it is expensive and difficult to scale across many prompts, models, and scoring dimensions. Recent learned or prompted judges, including VideoScore~\cite{he2024videoscore}, T2V-CompBench~\cite{sun2025t2v}, PhyWorldBench~\cite{gu2025phyworldbench}, and PhyAVBench~\cite{xie2025phyavbench}, provide more structured automated evaluation, but they remain limited to generic quality, visual physics, or audio-only sensitivity. In contrast, \agent is designed for joint audio-video physical commonsense, combining multimodal reasoning with deterministic audio-visual evidence to score each generation along five dimensions without requiring ground-truth recordings.


\section{\bench}
\label{sec:method}
\vspace{-3.5mm}
\begin{figure}[h]
  \vspace{1mm}
  \centering
  \includegraphics[width=0.88\linewidth]{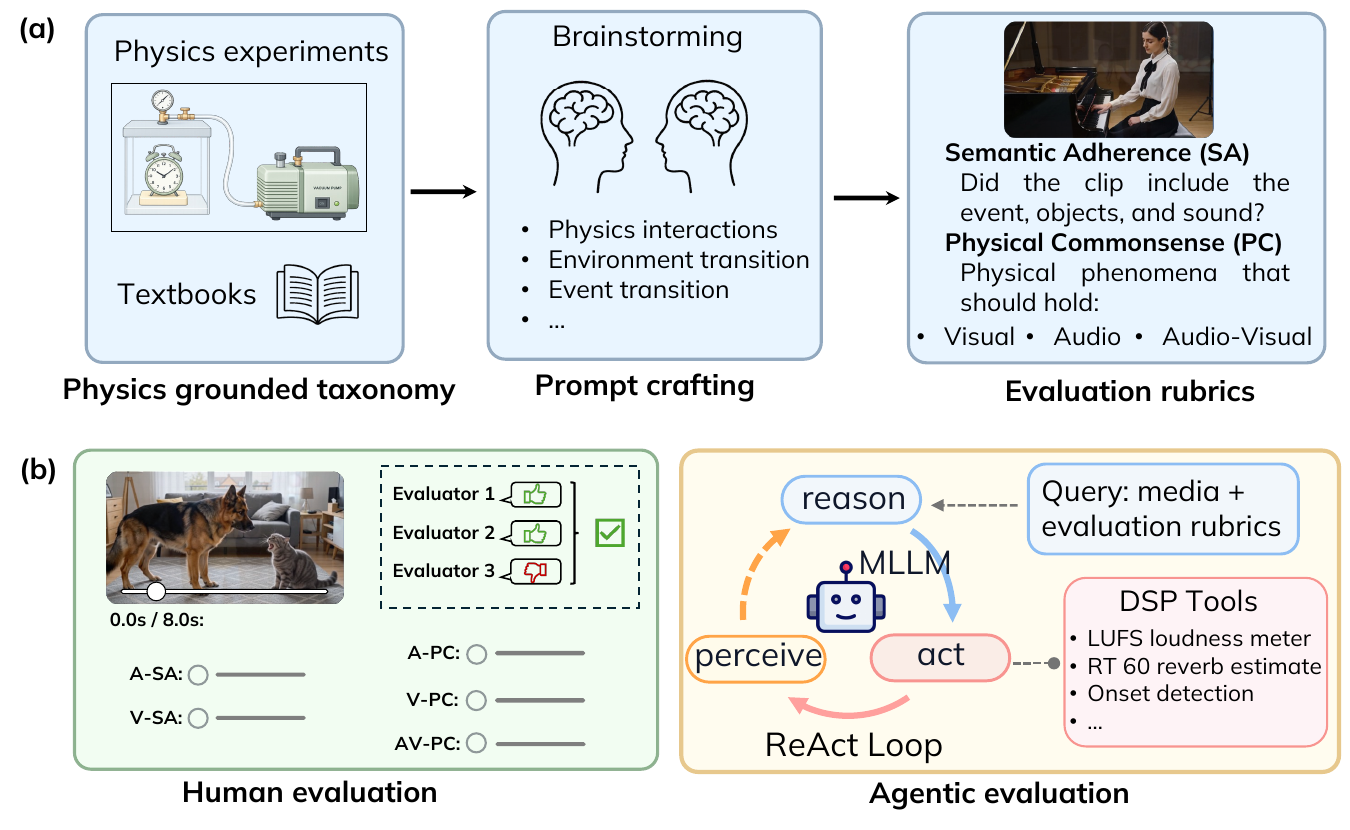}
  \vspace{-1.3mm}
  \caption{\textbf{\bench construction and evaluation pipeline.} (a)~A physics-grounded taxonomy organizes prompts by how the underlying physics evolve within a clip. Human-in-the-loop curation produces prompts that encode specific, verifiable acoustic outcomes, and each prompt is paired with a five-dimension evaluation rubric covering semantic adherence~(SA) and physical commonsense~(PC) across video (V), audio (A), and their cross-modal coupling (AV). (b)~Left panel: each generated clip is rated by three annotators from a pool of ten. Right panel: \agent, a pipeline that pairs a multimodal LLM model with deterministic audio-visual measurement tools.}
  \label{fig:overview}
\end{figure}

\bench evaluates whether joint audio-video generative models follow audio-visual physical commonsense rather than merely producing realistic-looking and plausible-sounding outputs. Prior benchmarks~\cite{gu2025phyworldbench,xie2025phyavbench} typically focus on a single modality, isolated events, or non-physical forms of audio-video alignment. By contrast, \bench is built around a \textit{Scene-Evolution Taxonomy} (Sec~\ref{subsec:taxonomy}) for joint audio-video physics, together with structured \textit{Evaluation Rubrics} (Sec~\ref{subsec:rubric}) for semantic and physical correctness within and across modalities, and a unified evaluation pipeline, \textit{Human and \agent Evaluation} (Sec~\ref{subsec:eval}), that combines human annotation with \agent-based automatic scoring. \bench contains 321 hand-written prompts (Appendix~\ref{app:curation}), including 268 physics-following prompts that cover 41 audio-visual physics principles and 53 Anti-AV-Physics prompts, with 2,763 rubric statements in total. Figure~\ref{fig:overview} illustrates the overall framework.

\subsection{Scene-Evolution Taxonomy}
\label{subsec:taxonomy}

\begin{figure}[h]
  \centering
  \includegraphics[width=1.0\linewidth]{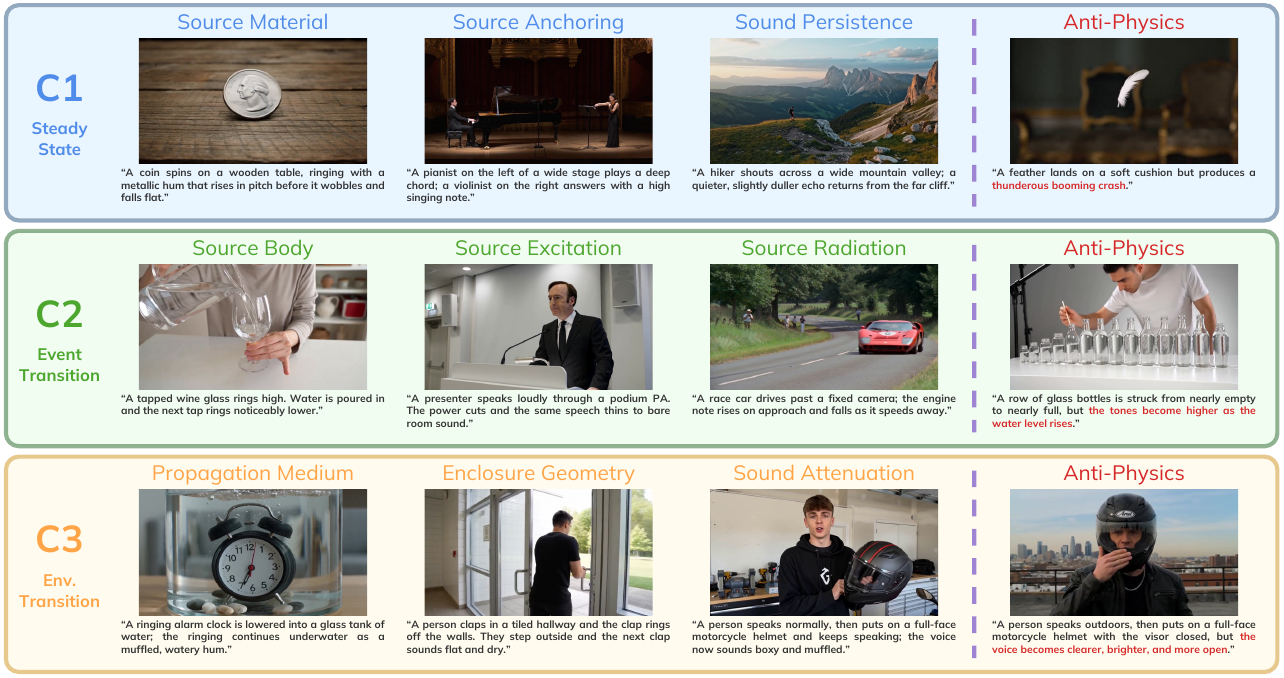}
  \vspace{-3.6mm}
  \caption{\textbf{Taxonomy overview of AV-Phys Bench.} Three scene categories of physics-following prompts, with a per-category Anti-AV-Physics subcategory in the rightmost column.}
  \label{fig:taxonomy}
\end{figure}

We organize the \bench prompts along a primary axis that captures how the underlying physics \emph{evolves} within a clip.
In the physical world, an audio-visual event is shaped by three factors: the source itself, the actions applied to the source, and the environments between source and listener. Our taxonomy turns these factors into three scene categories. In \emph{Steady State}, all three remain fixed. In \emph{Event Transition}, the action changes. In \emph{Environment Transition}, the environments change. The latter two introduce evaluation settings that no prior audio-visual benchmarks isolated (Table~\ref{tab:comparison}). Figure~\ref{fig:taxonomy} shows representative prompts and generations for each category. Each physics-following prompt has one scene category and one of the 41 physics principles in Table~\ref{tab:full_taxonomy}: the category specifies how the scene state evolves, and the principle specifies the physical principle that the generated clip should follow. Every clip is then scored on the same five dimensions (Section~\ref{subsec:rubric}).

\textbf{Category~1: Steady State.}
This category covers scenes whose source and environment remain unchanged over time, thus focusing on the intrinsic audio-visual correspondence. It includes three subcategories.
\emph{Source Material}: covers material-dependent timbre, spanning metal, wood, brittle, fluid, paper, and soft material classes.
\emph{Source Anchoring}: covers the relationship between audio and visible sources, including stereo lateralization, multi-source separation, and the diegetic versus non-diegetic distinction.
\emph{Sound Persistence}: covers the temporal evolution of sound in a fixed scene, including decay, echo, reverberation, and the propagation effects of a static medium or barrier.

\textbf{Category~2: Event Transition.}
This category introduces an action that changes the physical state of the source, testing whether the resulting audio-visual changes remain physically consistent. It includes three subcategories.
\emph{Source Body}: covers modifications to a vibrating element's length, tension, or mass loading that shift pitch.
\emph{Source Excitation}: covers variations in strike force, vocal effort, or gain control that shift loudness or timbre.
\emph{Source Radiation}: covers changes in source motion or resonance that shift pitch, loudness, or perceived location.

\textbf{Category~3: Environment Transition.} This category changes the propagation path between source and listener with the source held fixed, testing whether the corresponding audio-visual response remains physically consistent. It includes three subcategories. \emph{Propagation Medium}: covers transmission through vacuum, gases, liquids, and solids that modulate the perceived sound. \emph{Enclosure Geometry}: covers reflections from the surrounding space that produce reverberation, echo, or a dry response. \emph{Sound Attenuation}: covers material absorption, barrier obstruction, and ambient masking that reduce or muffle the perceived sound.

\textbf{Anti-AV-Physics subcategory.}
Within each scene category, an additional Anti-AV-Physics subcategory asks the model to render an outcome that deliberately violates the relevant physical principle while remaining faithful to the literal description (e.g., a cat opens its mouth and produces a perfectly lip-synced dog bark, breaking source-identity coupling while preserving causal and temporal coupling). Extending PhyWorldBench's anti-physics design~\cite{gu2025phyworldbench} to the cross-modal audio-visual setting, this subcategory probes whether models possess generative physics knowledge or merely encode physically consistent priors.

Together, these scene categories enable \bench to systematically evaluate whether joint audio-video generation models preserve physical consistency under diverse forms of scene evolutions.

\subsection{Evaluation Rubric}
\label{subsec:rubric}

To comprehensively evaluate physical commonsense in joint audio-video generation, the rubric must capture two distinctions. A model may fail semantically by omitting or misrendering the entities and sounds described in the prompt, or it may satisfy the prompt at a surface level while violating the underlying physics. These failures may occur within either modality or in the consistency between them. We therefore design the rubric around five dimensions that separate semantic adherence from physics aspects, both within each modality and across modalities.

Specifically, each generated clip is scored along five dimensions. For each dimension, we begin with a generic binary template and instantiate it into a prompt-specific yes/no statement. Four dimensions evaluate the two modalities separately: visual semantic adherence (\vsa), visual physical commonsense (\vpc), audio semantic adherence (\asa), and audio physical commonsense (\apc). The fifth, cross-modal physical commonsense (\avpc), evaluates whether the visual and audio streams remain consistent with the same underlying physical event.

\textbf{Within-modality dimensions.}
The four within-modality dimensions assess each stream on its own. \vsa and \asa measure whether the visual entities and sounds described by the prompt appear in the generated video and audio, respectively. \vpc evaluates whether the visual stream obeys real-world physics, including visible motion, contact, and material behavior. \apc evaluates whether the audio stream obeys real-world physics, including timbre, decay, propagation, and frequency content.

\textbf{Cross-modal physical commonsense.}
The fifth dimension, \avpc, evaluates whether visual and audio streams remain consistent with the same underlying physical event. This is the most distinctive aspect of our benchmark relative to prior audio-visual evaluations. In particular, \avpc captures four facets of cross-modal physical consistency: causal coupling, temporal coupling, spatial coupling, and source-identity coupling. Causal coupling requires the audible effect to follow its visible cause~\cite{owens2016visually,owens2018audio}. Temporal coupling requires the audio onset to align with the visible contact at the physically appropriate offset~\cite{tian2018audio,korbar2018cooperative}. Spatial coupling requires the sound to localize itself to its visible source~\cite{mo2023audio,chen2026scenebind}. Source-identity coupling requires the heard object to match the seen object~\cite{tian2021cyclic,zhao2018sound}. No prior audio-visual benchmark evaluates this dimension (Table~\ref{tab:comparison}).

\begin{figure}[t]
  \centering
  \includegraphics[width=1.0\linewidth]{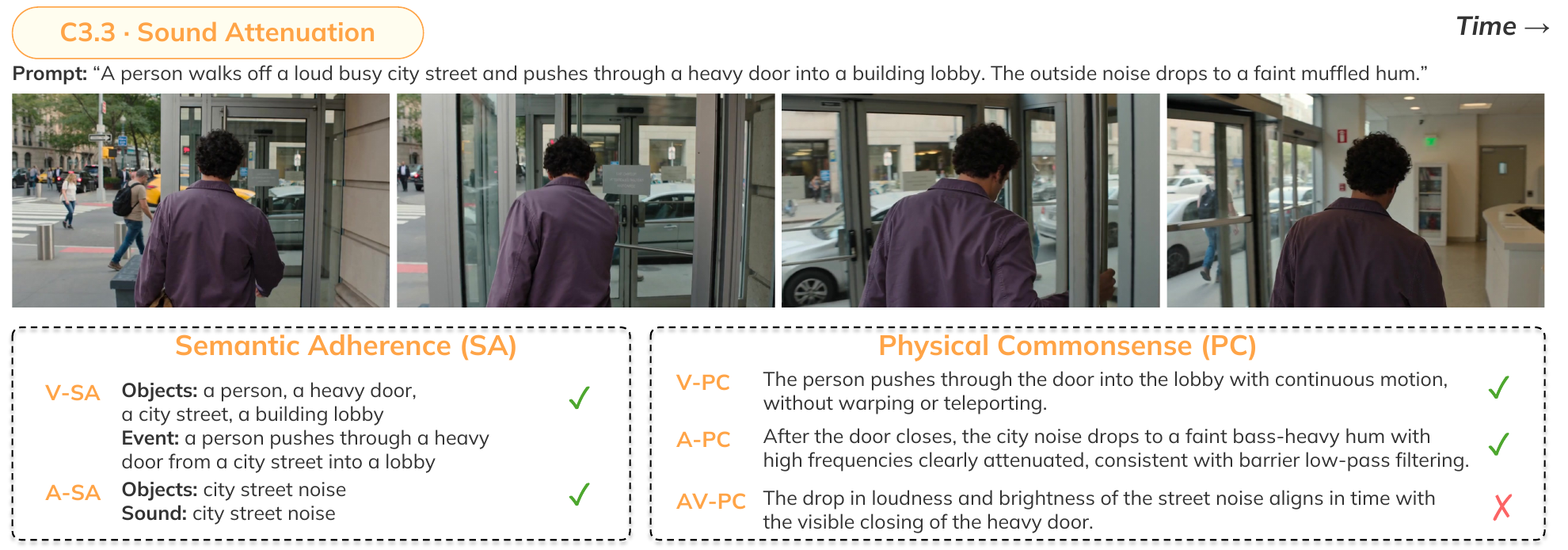}
  \caption{\textbf{Worked rubric example for a Category-3 Environment Transition prompt.} Top: four frames from the Seedance 2.0 generation showing the outdoor-to-indoor transition through a heavy door. Bottom: the prompt-specific rubric used for evaluation. \vsa and \asa check whether the described visual entities and sounds are present. \vpc, \apc, and \avpc check whether the generated video, audio, and their alignment obey the expected physical behavior. In this example, AV-PC tests whether the drop in loudness and high-frequency content occurs at the same time as the visible closing of the heavy door. Each statement receives a binary Y/N judgment.
  }
  \label{fig:rubric_example}
  \vspace{-5mm}
\end{figure}

\textbf{Worked example.}
To make the rubric concrete, we instantiate the five generic templates into prompt-specific binary Y/N statements for every prompt. Figure~\ref{fig:rubric_example} shows one such instance for a Category-3 Environment Transition prompt, where \avpc binds the audible reverb-to-dry transition to the visible threshold-crossing moment.

\textbf{Aggregation.}
A clip passes a given dimension only if every yes/no statement under that dimension receives a positive judgment. At the aggregate level, SA requires both \vsa and \asa, PC requires \vpc, \apc, and \avpc, and \emph{Both} requires both SA and PC. We adopt strict conjunction because physical consistency cannot be only partially satisfied: if the audio onset is mistimed relative to a visibly correct contact event, then the clip has failed cross-modal physics, even if its audio and visuals each appear plausible in isolation. Scoring each clip by the fraction of PC statements it passes gives the same order for the top five models (Appendix~\ref{app:bootstrap}).

\subsection{Human and \agent Evaluation}
\label{subsec:eval}

\textbf{Human evaluation.}
Human judgment provides the gold-standard reference for \bench. Each generated 8-second clip is evaluated using the same prompt-specific rubric described in Section~\ref{subsec:rubric}. Annotators view the clip together with its prompt and instantiated rubric statements, and answer every statement with a binary yes/no judgment. We use a pool of ten trained annotators, and each clip is rated independently by three annotators across all five dimensions. Across the seven evaluated models, this gives 2,247 clips and about 58,000 statement-level human judgments. Statement-level judgments are resolved by majority vote and then aggregated under the strict conjunction rule from Section~\ref{subsec:rubric}. This produces statement-level labels for semantic adherence, within-modality physical commonsense, and cross-modal physical consistency, rather than a single coarse overall score. Anti-AV-Physics clips use the same five dimensions and aggregation, but their PC statements check the outcome that the prompt requests, including the violation. For example, when a prompt asks for thunder before the lightning flash, its \avpc statements require the thunder to play first.

\textbf{\agent evaluation.}
To scale evaluation beyond the human panel, we introduce \emph{AV-Phys Agent}, a ReAct-style~\cite{yao2022react} agent that pairs Gemini~3.1~Pro Preview~\cite{comanici2025gemini} with deterministic measurement tools. Given a clip and its per-prompt rubric, \agent interleaves perception, reasoning, and tool calls: it watches and listens to the clip, decides what visual and acoustic quantities the rubric statements depend on, invokes the corresponding tools, and accumulates the returned measurements into its working description of the scene. Semantic statements such as \vsa and \asa are handled primarily through the model's own perception of the clip, while physics-sensitive statements such as \vpc, \apc, and \avpc are anchored to the relevant aduio DSP tools. The tools cover general digital signal processing methods including onset detection, pitch analysis, loudness metering, reverberation estimation, spectral comparison, stereo analysis, and silence checks. The resulting measurements are returned to the multimodal judge as structured evidence, which outputs a binary yes/no verdict for each rubric statement under a JSON schema. \agent is therefore a scalable evaluator tailored to the physical-commonsense structure of \bench. The complete tool inventory and the verbatim prompts are provided in Appendix~\ref{app:tool_inventory} and Appendix~\ref{app:agent_prompts}.

\section{Evaluation Results}
\label{sec:experiments}

We evaluate seven recent joint audio-visual generation models, including three proprietary models, Seedance 2.0~\cite{seedance2026seedance}, Kling 3.0 Omni~\cite{team2025kling}, and Veo 3.1~\cite{veo31}, and four open-source models, LTX-2.3~\cite{hacohen2026ltx}, Ovi~\cite{low2025ovi}, JavisDiT++~\cite{liu2026javisdit++}, and MagiHuman~\cite{chern2026speed}, on the constructed \bench. Model specifications and generation setup are detailed in Appendix~\ref{app:models}. Detailed qualitative analysis is provided in Appendices~\ref{app:qual_pc_failures} and~\ref{app:qual_anti_physics}.

\subsection{AV models show limited physical consistency compared to semantic adherence}
\label{subsec:overall}

First, we investigate the current models' generation abilities by five dimensions (V-SA, V-PC, A-SA, A-PC, AV-PC).  
From Table~\ref{tab:main_results}, we observe that model performances degraded consistently from semantic adherence (SA) to unimodal physical commonsense (PC) to cross-modal physical consistency (AV-PC) across all seven tested models. Among the seven models, we notice that even the strongest model (Seedance~2.0) dropped from 0.940 on \vsa to 0.750 on \avpc. Further, this pattern is more obvious for mid-tier models, e.g., Veo~3.1 (0.877 \vsa to 0.422 \avpc) and LTX-2.3 (0.519 \vsa to 0.239 \avpc). This finding is consistent with prior work~\cite{gu2025phyworldbench} showing that current T2AV models can produce semantically correct outputs without ensuring that either modality individually obeys physical laws. Beyond the within-modality physics gap, our fifth dimension \avpc further shows that current models struggle to align audio and video with the same underlying physical event.

\begin{wraptable}{r}{0.53\linewidth}
\vspace{0mm}
\centering
\setlength{\tabcolsep}{3pt}
\renewcommand{\arraystretch}{1.1}
\small
\begin{tabular}{@{}lccccc@{}}
\toprule
\textbf{Model} & \vsa & \asa & \vpc & \apc & \avpc \\
\midrule
\rowcolor{rowprop}\multicolumn{6}{l}{\emph{Proprietary Models}} \\
Seedance 2.0~\cite{seedance2026seedance}    & \textbf{0.940} & \textbf{0.933} & \textbf{0.840} & \textbf{0.769} & \textbf{0.750} \\
Kling 3.0 Omni~\cite{team2025kling}   & 0.925 & 0.840 & 0.716 & 0.489 & 0.556 \\
Veo 3.1~\cite{veo31}          & 0.877 & 0.854 & 0.519 & 0.425 & 0.422 \\
\midrule
\rowcolor{rowopen}\multicolumn{6}{l}{\emph{Open-source Models}} \\
LTX-2.3~\cite{hacohen2026ltx}          & \textbf{0.519} & \textbf{0.567} & \textbf{0.295} & \textbf{0.209} & \textbf{0.239} \\
Ovi 1.1~\cite{low2025ovi}      & 0.325 & 0.351 & 0.127 & 0.090 & 0.075 \\
JavisDiT++~\cite{liu2026javisdit++}       & 0.239 & 0.325 & 0.063 & 0.086 & 0.019 \\
MagiHuman~\cite{chern2026speed}      & 0.116 & 0.198 & 0.052 & 0.071 & 0.049 \\
\bottomrule
\end{tabular}
\caption{\textbf{Human evaluation: per-dimension scores on the 268 physics-following prompts across seven AV models.} Each score is the fraction of clips that pass the dimension. A clip passes only if the majority vote on every statement of that dimension is yes.}
\label{tab:main_results}
\vspace{-3mm}
\end{wraptable}
Furthermore, we note that the gap between proprietary and open-source models is substantial and consistent across all five dimensions. For example, the strongest open-source system, LTX-2.3, achieves 0.519 on \vsa and 0.239 on \avpc, compared to Veo~3.1's 0.877 and 0.422, respectively. Another notable inversion emerges in the open-source tier: \asa consistently exceeds \vsa (LTX-2.3: 0.567 vs.\ 0.519; Ovi 1.1: 0.351 vs.\ 0.325; JavisDiT++: 0.325 vs.\ 0.239), reversing the proprietary pattern in which \vsa always leads. This inversion suggests that current open-source and proprietary models exhibit different modality-level error profiles. Open-source models more often fail to show the described event, whereas proprietary models more often fail to produce the described sound (Appendix~\ref{app:sa_breakdown}).
 
\subsection{Per-Category Analysis}
\label{subsec:category}

\begin{table}[h]
\vspace{-3mm}
\centering
\scriptsize
\setlength{\tabcolsep}{3pt}
\renewcommand{\arraystretch}{1.0}
\resizebox{\textwidth}{!}{%
\begin{tabular}{l ccc ccc ccc ccc}
\toprule
\multirow{2}{*}{Model}
 & \multicolumn{3}{c}{C1: Steady State}
 & \multicolumn{3}{c}{C2: Event Transition}
 & \multicolumn{3}{c}{C3: Env. Transition}
 & \multicolumn{3}{c}{Overall}\\
\cmidrule(lr){2-4}\cmidrule(lr){5-7}\cmidrule(lr){8-10}\cmidrule(lr){11-13}
 & SA & PC & Both & SA & PC & Both & SA & PC & Both & SA & PC & Both \\
\midrule
\rowcolor{rowprop}\multicolumn{13}{l}{\emph{Proprietary Models}}\\
Seedance 2.0~\cite{seedance2026seedance}    & \textbf{0.906} & \textbf{0.732} & \textbf{0.717} & \textbf{0.886} & \textbf{0.552} & \textbf{0.552} & \textbf{0.944} & \textbf{0.722} & \textbf{0.722} & \textbf{0.903} & \textbf{0.660} & \textbf{0.653} \\
Kling 3.0 Omni~\cite{team2025kling}    & 0.835 & 0.433 & 0.417 & 0.752 & 0.267 & 0.267 & 0.861 & 0.250 & 0.250 & 0.806 & 0.343 & 0.336 \\
Veo 3.1~\cite{veo31}   & 0.787 & 0.315 & 0.307 & 0.819 & 0.124 & 0.124 & 0.750 & 0.139 & 0.139 & 0.795 & 0.216 & 0.213 \\
\midrule
\rowcolor{rowopen}\multicolumn{13}{l}{\emph{Open-source Models}}\\
LTX-2.3~\cite{hacohen2026ltx}        & \textbf{0.425} & \textbf{0.150} & \textbf{0.126} & \textbf{0.390} & \textbf{0.029} & \textbf{0.029} & \textbf{0.361} & \textbf{0.111} & \textbf{0.111} & \textbf{0.403} & \textbf{0.097} & \textbf{0.086} \\
Ovi 1.1~\cite{low2025ovi}      & 0.189 & 0.016 & 0.008 & 0.257 & 0.019 & 0.019 & 0.139 & 0.000 & 0.000 & 0.209 & 0.015 & 0.011 \\
JavisDiT++~\cite{liu2026javisdit++}     & 0.142 & 0.000 & 0.000 & 0.152 & 0.000 & 0.000 & 0.083 & 0.000 & 0.000 & 0.138 & 0.000 & 0.000 \\
MagiHuman~\cite{chern2026speed}     & 0.126 & 0.008 & 0.008 & 0.048 & 0.000 & 0.000 & 0.056 & 0.000 & 0.000 & 0.086 & 0.004 & 0.004 \\
\bottomrule
\end{tabular}%
}
\vspace{2mm}
\caption{\textbf{Human evaluation results across the three scene categories.} Per-prompt scores aggregated by strict conjunction over per-statement majority votes from three annotators ($n$=268 physics-following prompts). SA stands for visual and audio semantic adherence; PC stands for visual, audio, and cross-modal physical commonsense; Both stands for the intersection of SA and PC. SA, PC, and Both require the listed dimensions to pass on the same clip, so they are at or below the per-dimension scores in Table~\ref{tab:main_results}.}
\label{tab:human_eval_main}
\vspace{-5mm}
\end{table}
 
Having established the overall inequality (PC $<$ SA), we next investigate the detailed per-category evaluation results. Table~\ref{tab:human_eval_main} summarize them by scene category: Steady State (C1, $n{=}$127), Event Transition (C2, $n{=}$105), and Environment Transition (C3, $n{=}$36).
 
\textbf{Consistency across categories.}
The two findings from the prior section~\ref{subsec:overall} were reconfirmed within every category.
First, PC $<$ SA holds uniformly, e.g., Seedance~2.0 on C1 achieves SA~$=$~0.906 while PC~$=$~0.732; on C2, SA~$=$~0.886 while PC~$=$~0.552.
Second, the proprietary models again perform better than open-source models. Further, we observed that even Seedance~2.0's worst category-level Both (0.552 on C2) still exceeded LTX-2.3's best (0.126 on C1). This consistency across categories showed that the physics gaps were inherent structural, rather than prompt-caused artifacts. We further bootstrap the 268 prompts 2,000 times: the standard deviation of PC is at most 0.03, and the ranking of the top four models does not depend on which prompts are included (Appendix~\ref{app:bootstrap}). 
 
\textbf{Transition scenes are the difficulty frontier.}
More importantly, our \bench provides a novel perspective in quantifying physical scenes (from steady-state to dynamics). The key category-level observation was that Steady State was consistently easier than both transition categories: C1~$>$~C2~$\approx$~C3. For proprietary models, the drop from C1 to C2 is pronounced: Kling~3.0~Omni falls from 0.433 to 0.267, Veo~3.1 from 0.315 to 0.124, and even Seedance~2.0 from 0.732 to 0.552.
Event Transition is the hardest category for most proprietary models, while Environment Transition poses comparable difficulty.
The open-source tier amplifies this pattern: LTX-2.3 drops from PC~$=$~0.150 on C1 to 0.029 on C2; Ovi, JavisDiT++, and MagiHuman each score 0.000 on at least one transition category. Our evaluation results was the first systematic measurement of the gap between sustaining a static physical configuration and tracking dynamical transition. Static physical plausibility is a fundamentally easier task than following a dynamical chain in which a visible action must produce a specific acoustic consequence. Thus, the three categories proposed in \bench help to isolate the explicit generative frontier where current models break down.

\subsection{Agent Evaluation}
\label{subsec:agent}

\begin{table}[t]
\centering
\small
\setlength{\tabcolsep}{8pt}
\renewcommand{\arraystretch}{1.1}
\begin{tabular}{l cccccc}
\toprule
\textbf{Method} & \vsa & \asa & \vpc & \apc & \avpc & \textbf{Avg. $\pm$ std} \\
\midrule
MLLM-as-judge baseline & 0.803          & 0.724          & 0.753          & 0.616          & 0.685          & 0.716 $\pm$ 0.071 \\
\agent                 & \textbf{0.830} & \textbf{0.770} & \textbf{0.797} & \textbf{0.768} & \textbf{0.768} & \textbf{0.787 $\pm$ 0.027} \\
\bottomrule
\end{tabular}
\vspace{2mm}
\caption{\textbf{Per-dimension agreement with human-majority labels.} Agreement is measured per (model, prompt, dimension) cell after strict-AND aggregation from per-statement majority votes. Avg. $\pm$ std is the sample mean and standard deviation across the five rubric dimensions.}
\label{tab:human_consistency}
\vspace{-5mm}
\end{table}

We validate \agent against the human-majority labels, which serve as the standard reference for \bench. As a baseline, we compare against an MLLM-as-a-judge evaluator without tool grounding, following the general evaluation paradigm used in prior work~\cite{gu2025phyworldbench, meng2024towards, bansal2025videophy}. Across all (model, dimension) cells that drive the leaderboard in Table~\ref{tab:human_eval_main}, \agent achieves an overall Pearson correlation of $r{=}0.934$ with human-majority labels, compared to $r{=}0.890$ for the MLLM-as-judge baseline (full per-dimension breakdown in Appendix~\ref{app:agent_correlation}). The improvement is concentrated on the audio side, where deterministic acoustic measurements add evidence beyond the model's native perception: \asa ($r{=}0.988$ vs.\ $0.883$) and \apc ($r{=}0.967$ vs.\ $0.909$).

Table~\ref{tab:human_consistency} reports per-dimension agreement with human-majority labels, i.e., binary accuracy against the majority human judgment for each dimension. \agent achieves an average agreement of 0.787, compared to 0.716 for the baseline, and outperforms it on all five dimensions. The largest gains appear on \apc ($+0.152$) and \avpc ($+0.083$). These results highlight the importance of tool-grounded evaluation for physics-sensitive judgments, where multimodal perception alone is less aligned with human annotations.

We further ablate the tool configuration using three variants: audio-DSP tools only, visual tools only (support frame extraction and zoomed-crop inspection), and the full audio-visual tool set. Audio-DSP tools (\agent setting) account for most of the improvement over the MLLM baseline, while visual tools alone add less (average agreement 0.744). One possible explanation is that the underlying multimodal LLM is already relatively strong at visual interpretation, whereas the audio and cross-modal physics dimensions benefit more directly from explicit measurements. Full ablation results, results on an open-source backbone, and a ranking-fidelity analysis based on Pearson correlation between evaluator and human pass rates were reported in Appendix~\ref{app:agent_tool_ablation} and Appendix~\ref{app:agent_correlation}.

\newpage
\subsection{Anti-AV-Physics}
\label{subsec:anti}

\begin{wraptable}{r}{0.43\linewidth}
\vspace{-4.4mm}
\centering
\setlength{\tabcolsep}{4pt}
\renewcommand{\arraystretch}{1.1}
\small
\begin{tabular}{@{}lccc@{}}
\toprule
\textbf{Model} & Phys. & Anti & Drop(\%) \\
\midrule
Seedance 2.0~\cite{seedance2026seedance}    & 0.660 & 0.208 & 68.5\% \\
Kling 3.0 Omni~\cite{team2025kling}   & 0.343 & 0.189 & 44.9\% \\
Veo 3.1~\cite{veo31}          & 0.216 & 0.113 & 47.7\% \\
\bottomrule
\end{tabular}
\caption{\textbf{Anti-AV-Physics.} A high PC means the model successfully produced the requested violation rather than defaulting to physics. Drop \% $= \frac{(\text{Phys.}-\text{Anti})}{\text{Phys.} }\times 100\%$ is the relative collapse from physics-following to anti-physics prompts.}
\label{tab:anti_physics}
\vspace{-3mm}
\end{wraptable}
Anti-AV-Physics prompts require models to follow the literal description while violating the underlying physical principle. As the open-sourced models have limited PC scores, we focus on 
proprietary models. Table~\ref{tab:anti_physics} summarizes PC scores on physics-following prompts ($n{=}$268) with anti-physics prompts ($n{=}$53). We observed that all proprietary models collapsed sharply (Drop: Seedance 68.5\%; Kling 44.9\%; Veo 47.7\%). A high Anti score means that a model can follow an instruction that contradicts physics, as counterfactual and creative generation require. The low scores point to different causes across models. Seedance~2.0 has relatively stronger physical priors, so when asked to violate physics it follows those priors and lacks the instruction-following ability to override them (Appendix~\ref{app:qual_anti_physics}). Kling~3.0~Omni and Veo~3.1 fail for a combination of reasons, as their physical priors are weaker and they do not follow the instruction either. For open-source models, semantic adherence is the current bottleneck, so their Anti scores do not yet reflect physical priors or instruction following. A regeneration of 195 prompts with Seedance~2.0 suggests that removing the explicit physical outcome from the prompt has little effect on the results (PC 0.626 vs.\ 0.595, $p=0.49$; Appendix~\ref{app:implicit}).

\section{Conclusion}
\label{sec:conclusion}
\vspace{-2mm}
We introduce \bench, the first benchmark for evaluating physical commonsense in joint audio-video generation. \bench combines a scene-evolution taxonomy with physics-grounded prompts and per-prompt rubrics to define the semantic and physical expectations for each clip. We pair prompt-specific human evaluation with \agent, a scalable evaluator that uses multimodal reasoning grounded in deterministic audio-visual measurement tools. Together, they assess whether the generated video and audio satisfy these physics-grounded expectations. Finally, we described the broad impact, limitations, and future work in the appendix~\ref{sec:impact}.

\begin{ack}
This work was supported in part by an Amazon Research Award. The authors thank High Performance Computing at The University of Texas at Dallas (HPC@UTD) for providing computing resources and support. Any opinions, findings, and conclusions expressed in this material are those of the authors and do not necessarily reflect the views of Amazon or HPC@UTD.
\end{ack}

\newpage
\bibliographystyle{plain}
\bibliography{references}

\newpage
\appendix
\addtocontents{toc}{\protect\setcounter{tocdepth}{1}}
\renewcommand{\contentsname}{Appendix Contents}
{\hypersetup{hidelinks}\tableofcontents}

\newpage
\section{Broader Impact and Limitations}
\label{sec:impact}
 
\textbf{Broader impact.}
\bench provides the first systematic diagnostic for where joint audio-video models fail on physics, offering model developers a concrete optimization target, i.e., the causal interaction between visible actions and acoustic consequences that Event Transition and Environment Transition isolate. Second,
beyond guiding architecture and training improvements for T2AV models, the taxonomy and rubric are input-modality agnostic. Thus, they can extend naturally to image-to-audio-video~\cite{veo31,seedance2026seedance}, video-to-audio~\cite{su2024vision,liu2024tell,pian2026omnisonic,dai2026omni2sound}, and audio-visual editing pipelines~\citep{xing2024survey}. Third, as generative models are increasingly positioned as world simulators~\cite{bruce2024genie}, physical realism across modalities becomes a prerequisite. Our \bench offers a concrete test of this capability. Finally,
our \agent pipelines actually constitute verifiable reward signals that could support reinforcement learning from verifiable rewards (RLVR) for physics-grounded post-training of joint audio-visual generative models~\citep{xue2026systematic, li2026video}.

\textbf{Limitations and future work.}
Our work has a few limitations. First, all prompts are designed in English and target eight-second clips. Extending our \bench to include multilingual prompts~\cite{zhang2024virbo} and longer-duration videos will be the next step. Second, our binary Y/N rubric trades severity information for annotator reliability. Thus, ordinal scales are a natural next step for better aligning human preferences. Last, while our ReACT agent workflow is actually MLLM model agnostic, we only test two backbones: the closed-source Gemini~3.1~Pro Preview~\cite{comanici2025gemini} and the open-source Qwen3-Omni-30B-A3B-Instruct~\cite{xu2025qwen3omni}. Testing with additional open-source/closed-source multimodal models (e.g., Qwen3.5-Omni~\cite{team2026qwen3}, GPT5~\citep{singh2025openai}) and expanding the human evaluation panel would greatly strengthen both the generalizability and the representativeness of the evaluation on \bench.

\section{Data and Code Availability}
\label{app:availability}

All prompts, rubrics, taxonomy specifications and evaluation results are released at \url{https://huggingface.co/datasets/ZijunCui/AV-Phys-Bench}, with a small public sample at \url{https://huggingface.co/datasets/ZijunCui/AV-Phys-Bench-Sample}. The 195 revised prompts and their generated clips are released as an extension of the benchmark.

The MLLM-as-judge baseline and \agent implementation are released at \url{https://github.com/ZijunCui02/AV-Phys}.
\section{Model Specifications and Generation Setup}
\label{app:models}

We evaluate seven recent joint audio-video generation models on \bench. All clips are 8-second, 16:9 (or near-16:9) videos with synchronized native audio at the providers' default sampling and guidance settings, with one exception noted below.

\textbf{Commercial / proprietary generators (API).}
We access these models only through their official APIs, with audio generation enabled and the providers' default scheduler and guidance.
\begin{itemize}[leftmargin=*, itemsep=0pt]
\item \textbf{Seedance 2.0}: Volcengine Ark API, model \texttt{doubao-seedance-2-0-260128}; 720p (1280$\times$720), 16:9, 24\,fps, 8\,s, audio on.
\item \textbf{Kling 3.0 Omni}: Kling API, \texttt{kling-v3-omni} Pro tier; 1080p (1920$\times$1080), 16:9, 24\,fps, 8\,s, audio on.
\item \textbf{Veo 3.1}: Google Vertex AI, \texttt{veo-3.1-fast-generate-001}; 1080p (1920$\times$1080), 16:9, 24\,fps, 8\,s, audio on.
\end{itemize}

\textbf{Open-source generators (code + weights).}
For each open-source model we run inference from the official public repository at the released checkpoint, using the recommended sampling configuration. All four are released under Apache-2.0.
\begin{itemize}[leftmargin=*, itemsep=0pt]
\item \textbf{LTX-2.3} (Apache-2.0): \url{https://github.com/Lightricks/LTX-Video}; 1920$\times$1088, 24\,fps, $\sim$8\,s, audio on.
\item \textbf{Ovi 1.1} (Apache-2.0): \url{https://github.com/character-ai/Ovi}; \texttt{960x960\_10s} checkpoint, 24\,fps. Ovi 1.1 is the only model whose native checkpoint outputs 10\,s clips instead of 8\,s, and this is a fixed model-side setting that does not expose an 8\,s configuration; we therefore evaluate Ovi at its native 10\,s duration.
\item \textbf{JavisDiT++} (Apache-2.0): \url{https://github.com/JavisDiT/JavisDiT}; 854$\times$480, 16\,fps, 5.0625\,s (81 frames) native output with 16\,kHz audio. We pad each clip to 8\,s with a frozen final frame and silent tail so all evaluators view the same nominal duration across models.
\item \textbf{MagiHuman} (Apache-2.0): \url{https://github.com/GAIR-NLP/daVinci-MagiHuman}; 1080p super-resolution pipeline (\texttt{sr\_1080p}), 1920$\times$1088, 25\,fps, 8\,s.
\end{itemize}

\section{Qualitative Examples: Semantic Adherence with Physical Failure}
\label{app:qual_pc_failures}

This appendix complements the headline finding that current joint audio-video models are far stronger at semantic adherence than at physical commonsense. We show clips that humans unanimously rated as semantically adherent (every \vsa and \asa statement passes) but on which at least one physical-commonsense statement fails. We surface cases whose physical failure is on the audio side (\apc) or in the cross-modal interaction (\avpc), because these are the dimensions where the headline gap is largest. Figures are ordered to match the model order in Table~\ref{tab:human_eval_main}.

\begin{figure}[h]
\centering
\includegraphics[width=\linewidth]{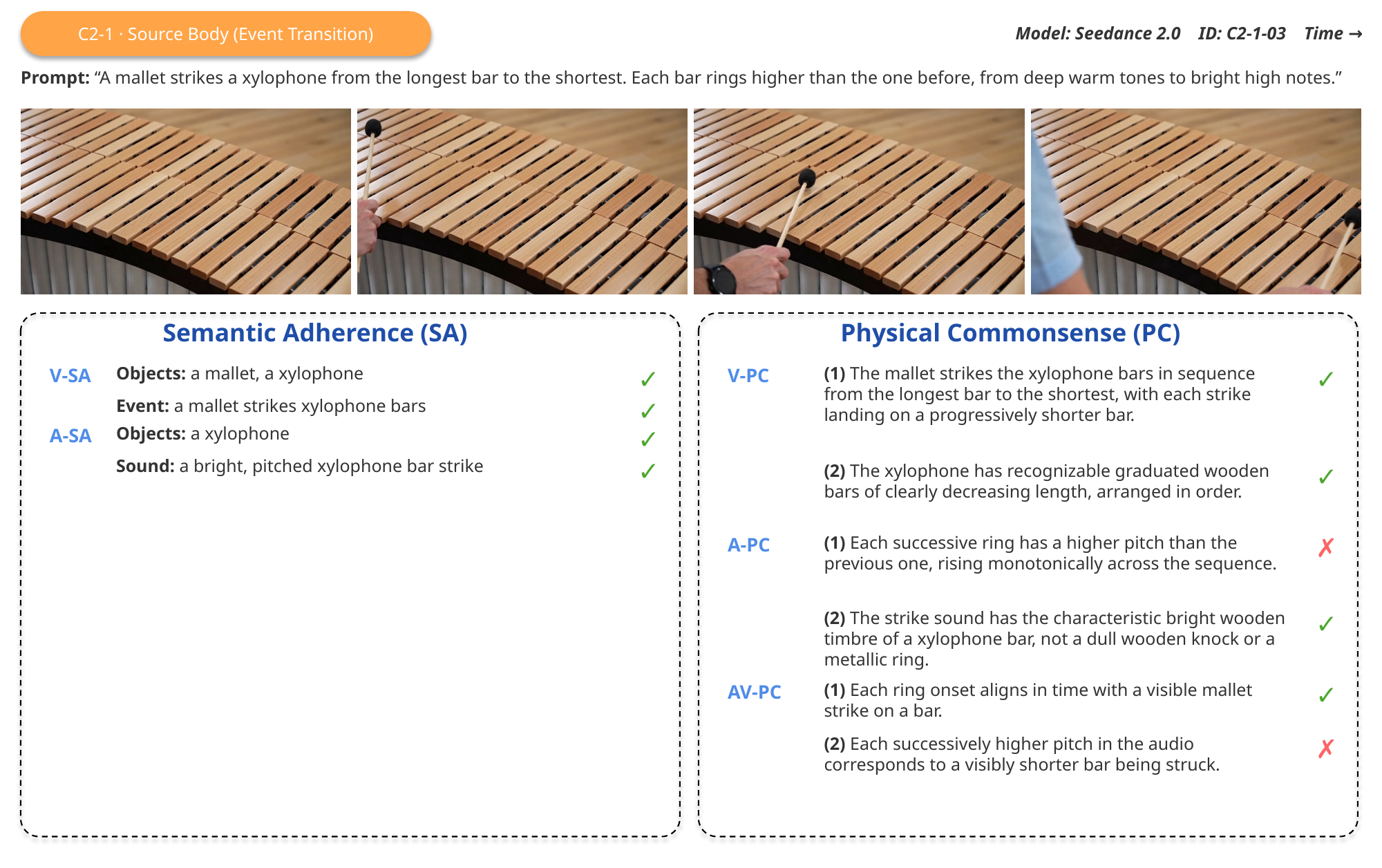}
\caption{Seedance 2.0, Event Transition. The clip strikes the xylophone bars from the longest to the shortest, matching the visible event the prompt specifies. The generated audio plays a recognisable melody whose pitch contour rises and falls instead of rising monotonically with each successive shorter bar, so the audio fails the size-to-pitch rule the prompt encodes.}
\label{fig:qual_pc_seedance}
\end{figure}

\begin{figure}[h]
\centering
\includegraphics[width=\linewidth]{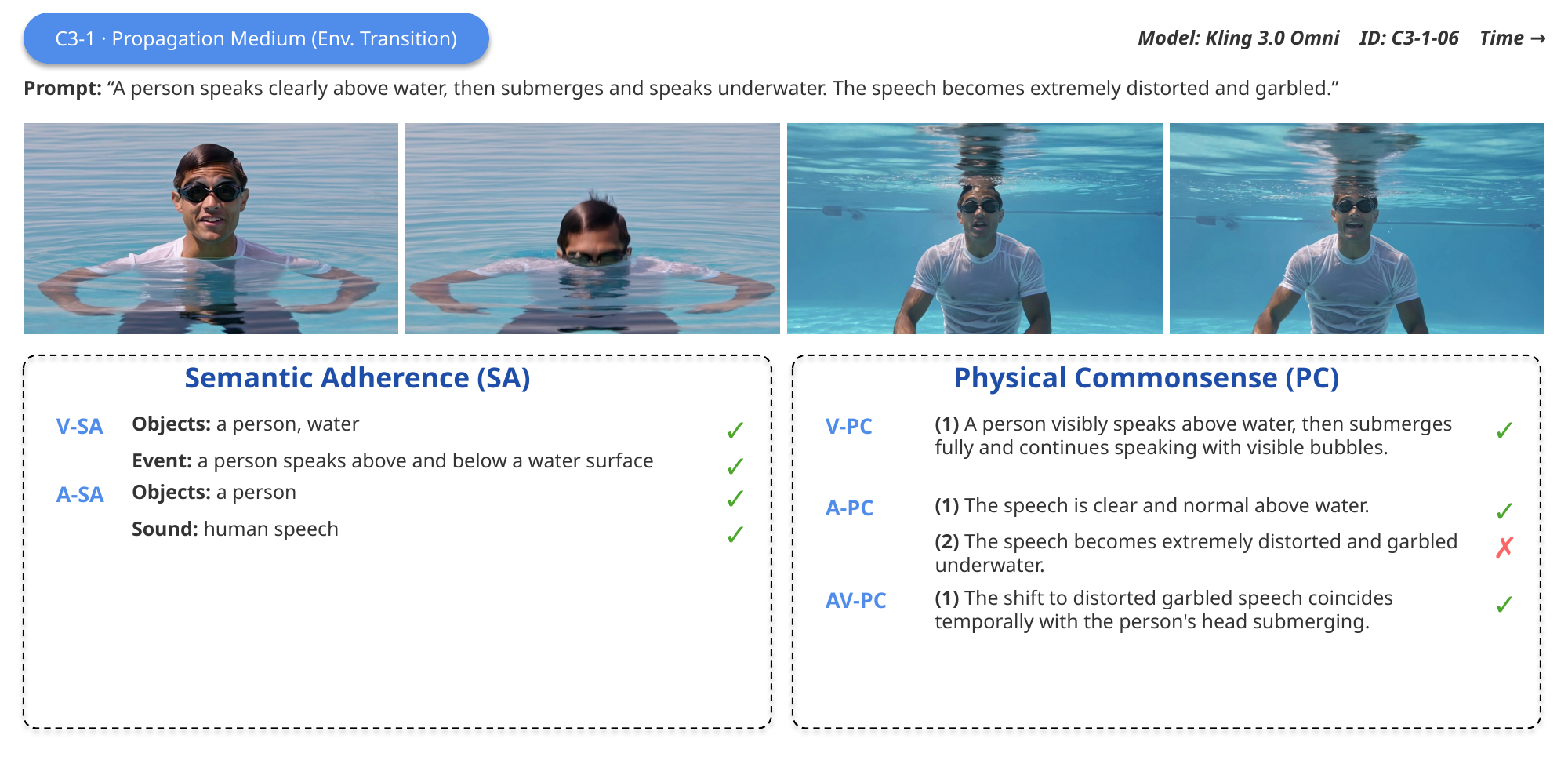}
\caption{Kling 3.0 Omni, Environment Transition. The clip captures the speaker submerging mid-sentence, and the audio dips and slightly compresses at that moment, signalling that the model registers a state change. The submerged speech remains nearly fully intelligible with light reverberation rather than the heavily attenuated and distorted character expected when sound travels through water; the result reads as a generic audio filter applied at the transition rather than a coherent simulation of liquid as the propagation medium.}
\label{fig:qual_pc_kling}
\end{figure}

\begin{figure}[h]
\centering
\includegraphics[width=\linewidth]{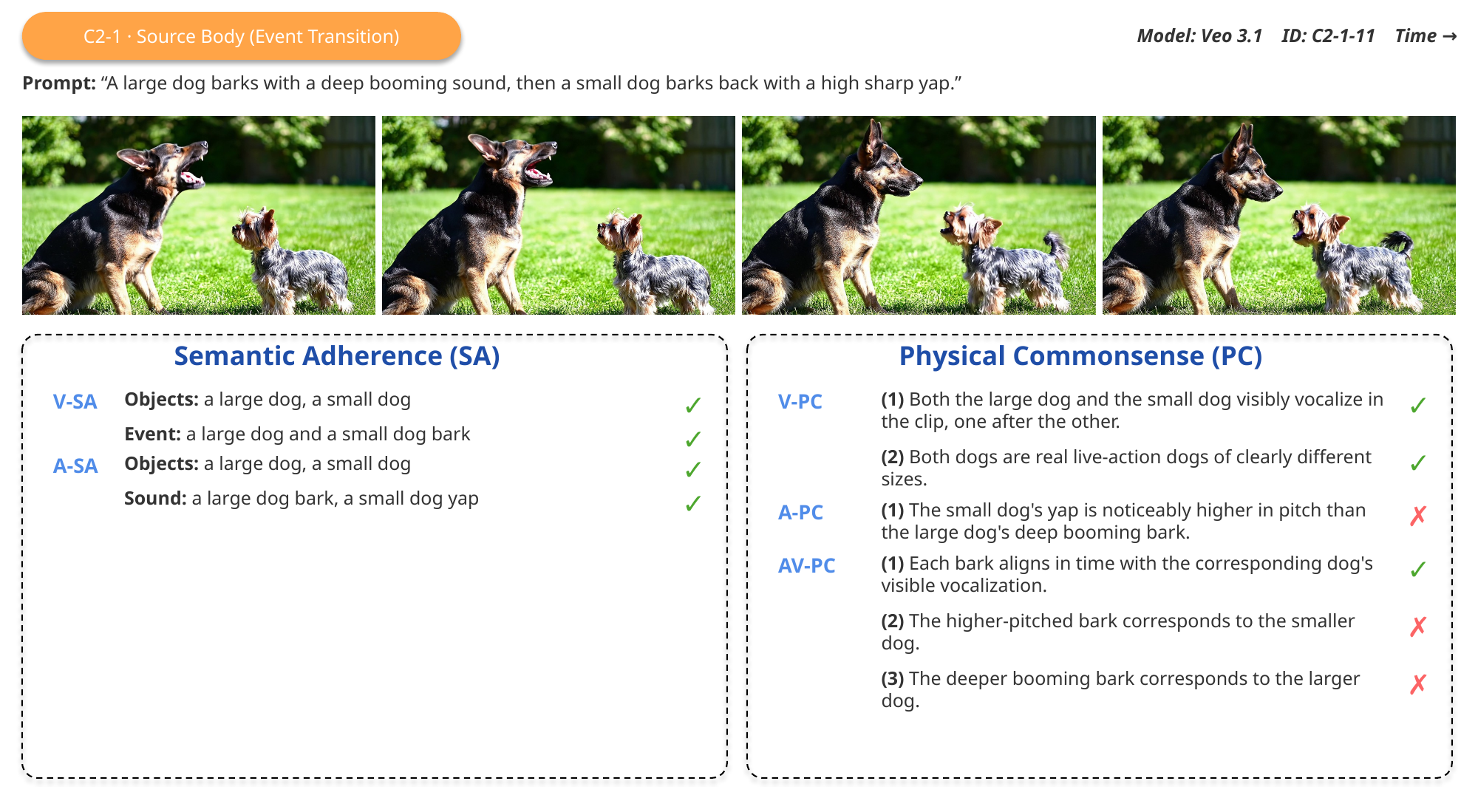}
\caption{Veo 3.1, Event Transition. The clip places a large dog on the left and a small dog on the right, with each bark visually synchronised to the corresponding animal. Both barks share the same deep, low-pitched timbre of a large dog, so the small dog's bark is acoustically indistinguishable from the large dog's, violating the size-to-pitch correspondence the prompt requires.}
\label{fig:qual_pc_veo}
\end{figure}

\begin{figure}[h]
\centering
\includegraphics[width=\linewidth]{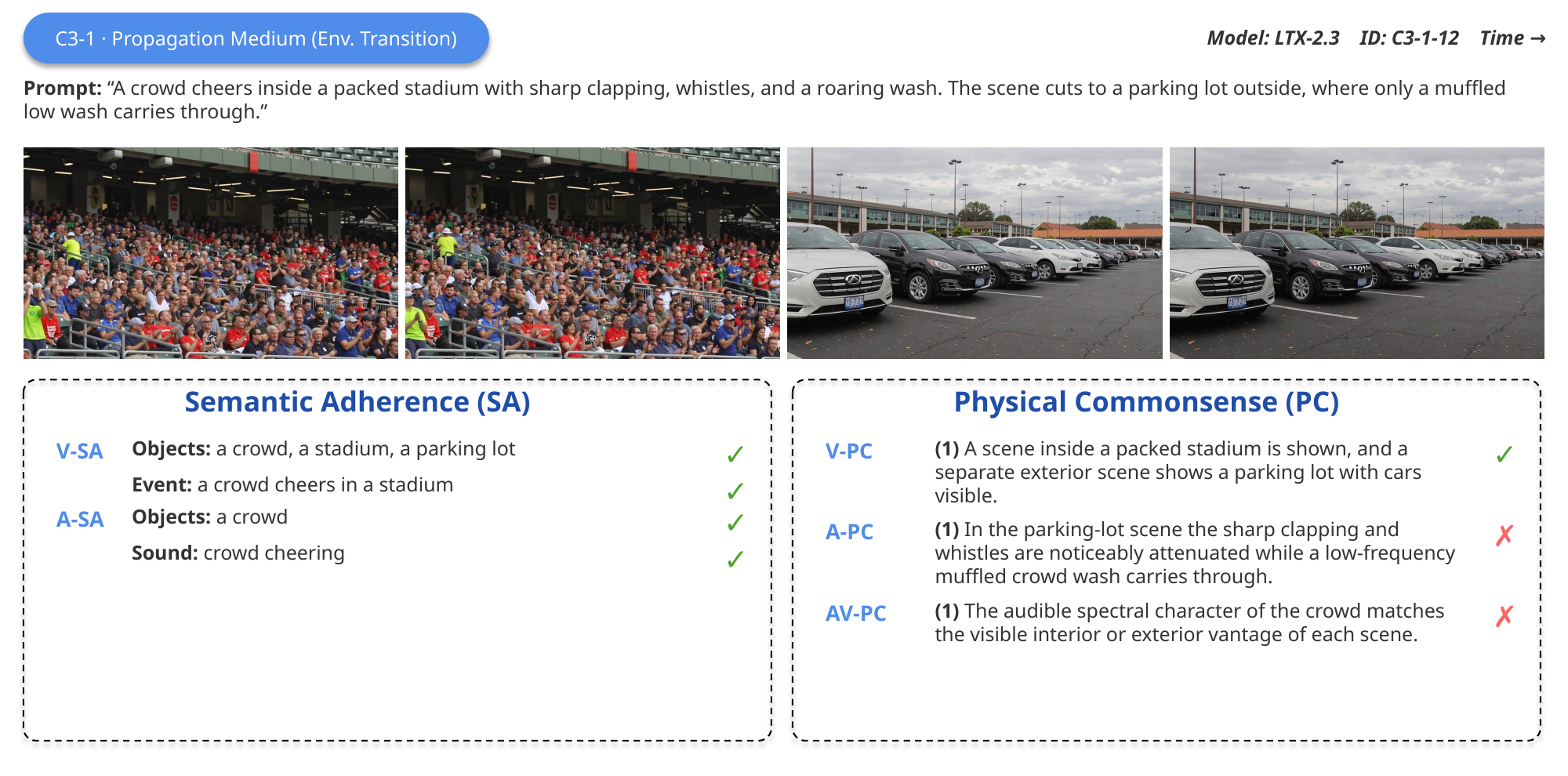}
\caption{LTX-2.3, Environment Transition. The clip cuts from a packed stadium interior to an exterior parking lot. The interior renders a moderate crowd applause that is roughly plausible, but the parking-lot scene drops the stadium ambience entirely instead of carrying it through as a high-frequency-attenuated remnant; the model treats the cut as a hard visual transition rather than a continuous acoustic environment with frequency-dependent decay.}
\label{fig:qual_pc_ltx}
\end{figure}

\begin{figure}[h]
\centering
\includegraphics[width=\linewidth]{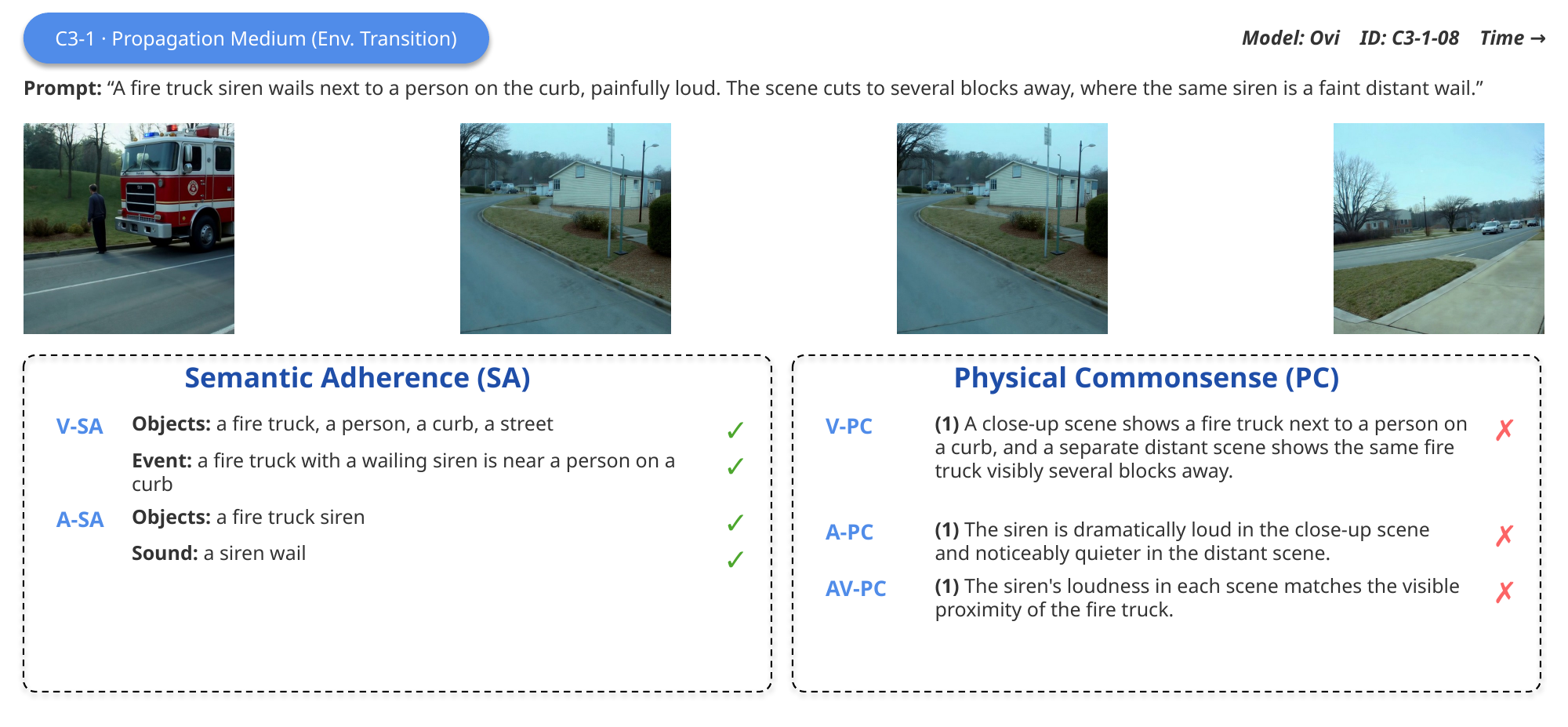}
\caption{Ovi, Environment Transition. The fire truck is visible only in the first frame and then disappears from the visual stream, so the prompt's near-versus-far transition is not depicted. The siren remains audible at constant loudness throughout, with no sign of the inverse-square decay that the same source receding several blocks should produce.}
\label{fig:qual_pc_ovi}
\end{figure}

\begin{figure}[h]
\centering
\includegraphics[width=\linewidth]{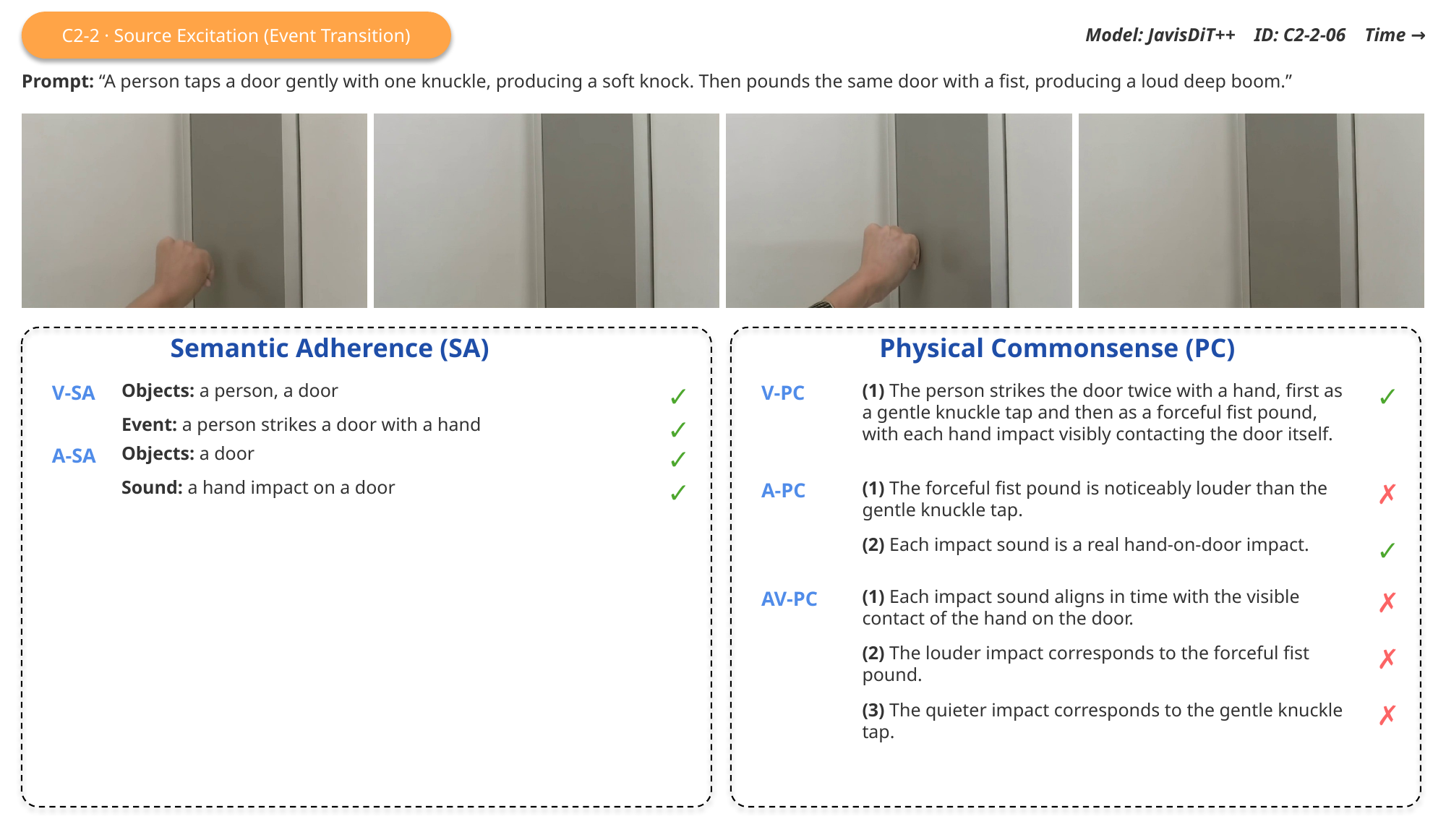}
\caption{JavisDiT++, Event Transition. The clip shows a person knocking gently three times and then striking the door once with visibly more force, so the visual stream contains the requested change in strike strength. The four knock sounds are nearly identical in volume and tone, and the audio onsets do not consistently align with the visible contacts, compounding the strike-force-to-loudness failure with a cross-modal sync issue.}
\label{fig:qual_pc_javisdit}
\end{figure}

\begin{figure}[h]
\centering
\includegraphics[width=\linewidth]{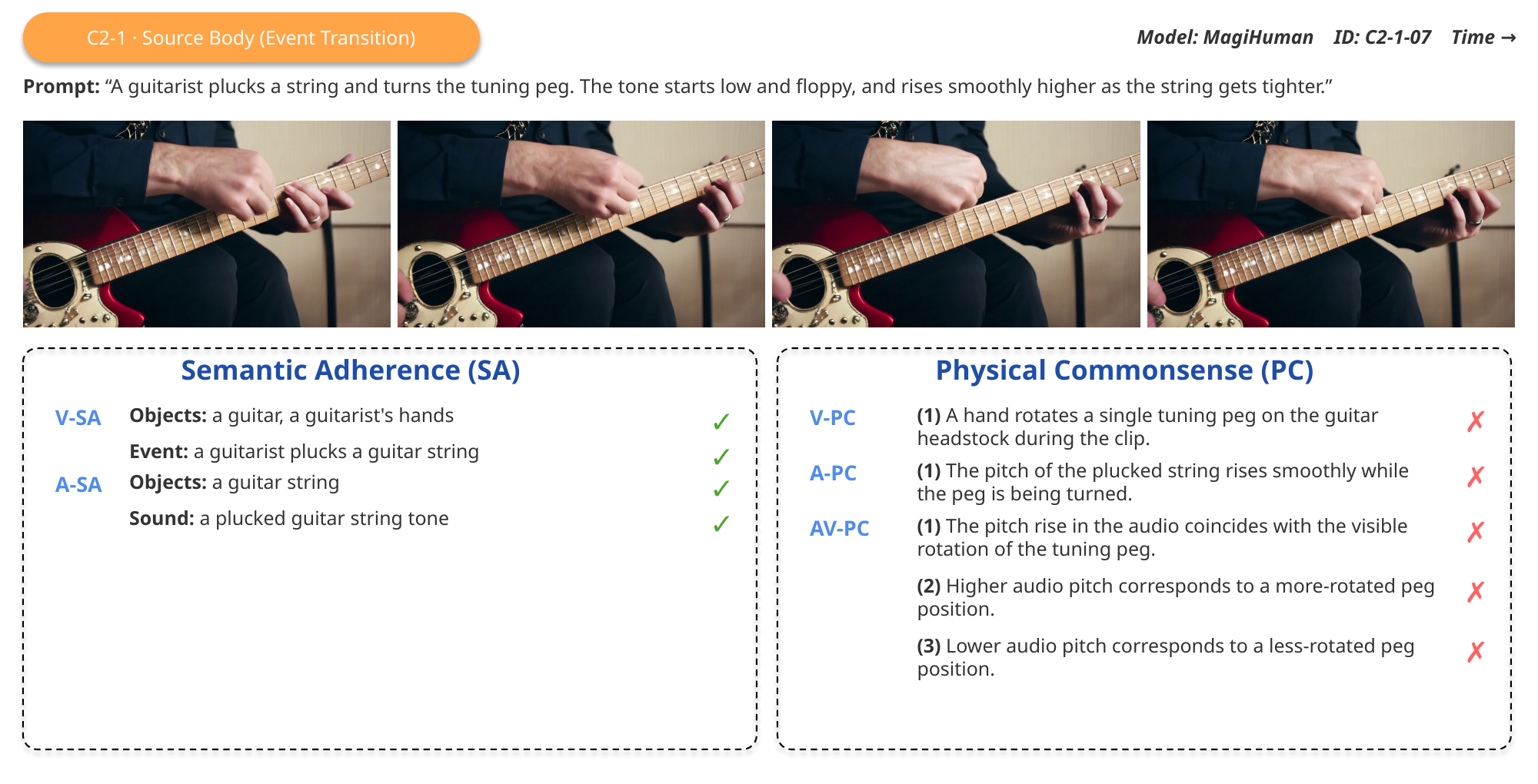}
\caption{MagiHuman, Event Transition. The clip shows a guitarist plucking a string but never visibly turning the tuning peg, so the action that the prompt makes the source of the pitch change is absent from the visual stream. We do not retune the prompt to fit MagiHuman's behaviour, since the same prompt is fed to every model for fairness; this drop is consistent with the model's tendency to paraphrase the prompt rather than execute fine-grained details.}
\label{fig:qual_pc_magihuman}
\end{figure}

\clearpage

\section{Qualitative Examples: Anti-AV-Physics}
\label{app:qual_anti_physics}

This appendix complements the Anti-AV-Physics finding: when prompts deliberately ask for an audio-visual outcome that violates real-world physics, current models predominantly default to physically plausible outputs and therefore fail the anti-prompt's intended target. We show one Seedance 2.0 clip per scene category. Even the strongest proprietary model satisfies the prompt's semantic surface (\vsa, \asa pass) but fails to render the physics-violating consequence the rubric tests, illustrating that the model encodes physically consistent priors rather than freely composing the requested cross-modal violation.

\begin{figure}[h]
\centering
\includegraphics[width=\linewidth]{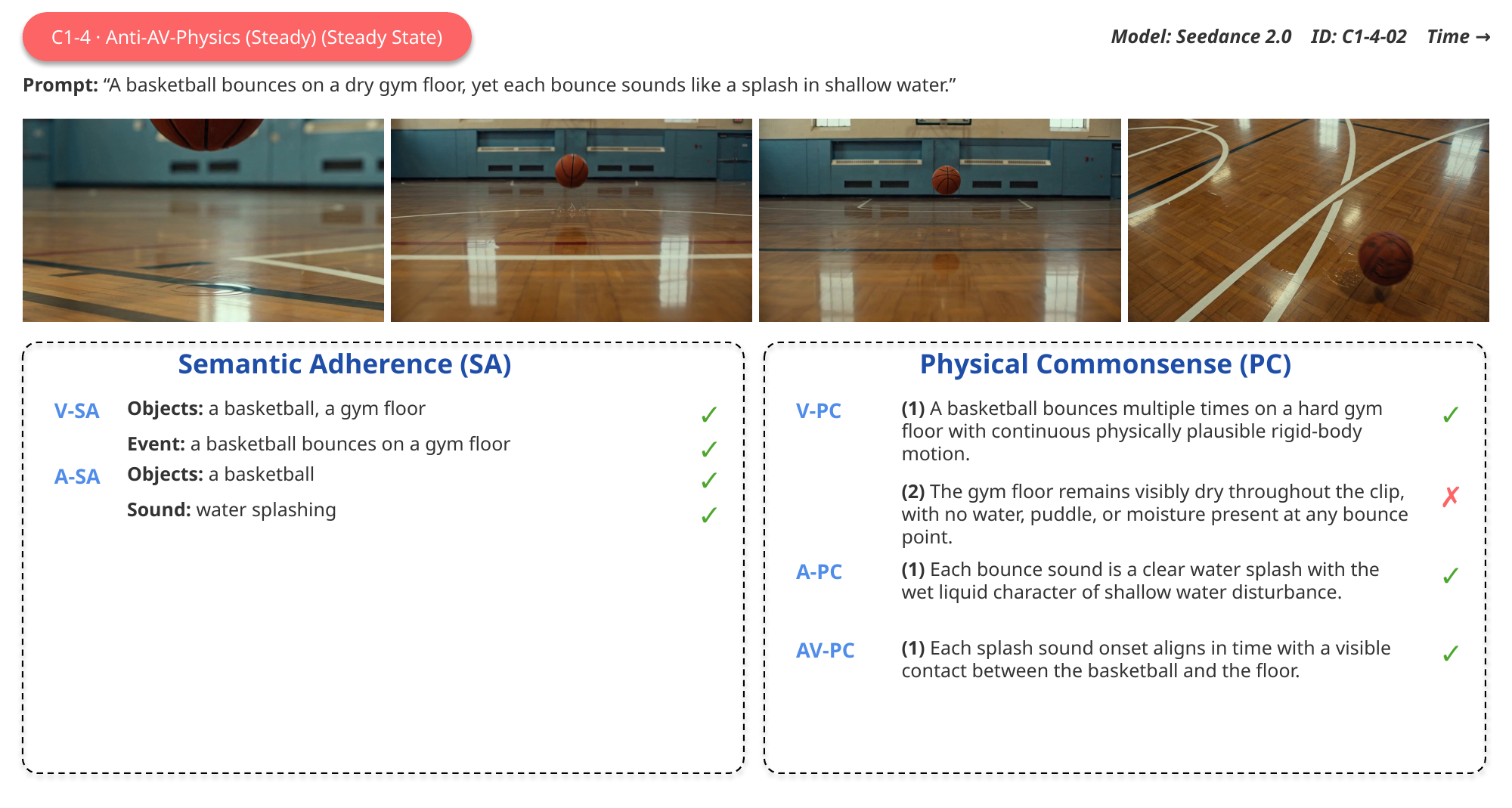}
\caption{Seedance 2.0, Steady State Anti-Physics. The model paints the gym floor with a thin layer of water so each bounce visibly kicks up a small splash, and the audio mixes a splash sound with the bounce sound to fit that visual. The clip sidesteps the prompt's deliberate dry-floor / splash-sound mismatch by editing the visible scene to make the splash physically plausible rather than rendering the requested cross-modal violation.}
\label{fig:qual_anti_c1}
\end{figure}

\begin{figure}[h]
\centering
\includegraphics[width=\linewidth]{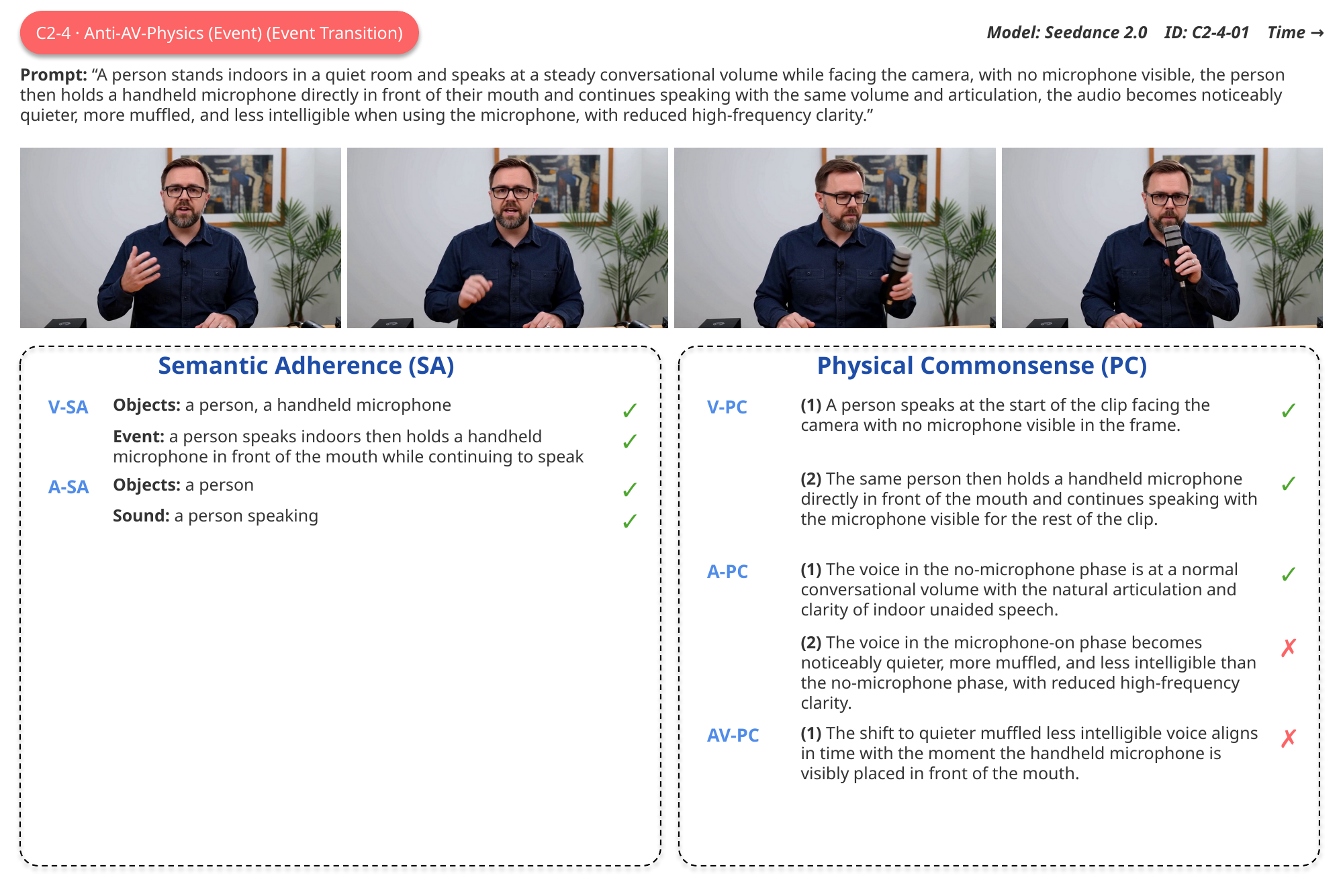}
\caption{Seedance 2.0, Event Transition Anti-Physics. With the handheld microphone held close to the mouth, the audio adds typical proximity-effect plosives and breath bursts of real handheld-microphone use, rather than the quieter and more muffled voice the anti-prompt asks for. The model executes the standard microphone-as-amplifier physics instead of the inverted outcome the rubric tests.}
\label{fig:qual_anti_c2}
\end{figure}

\begin{figure}[h]
\centering
\includegraphics[width=\linewidth]{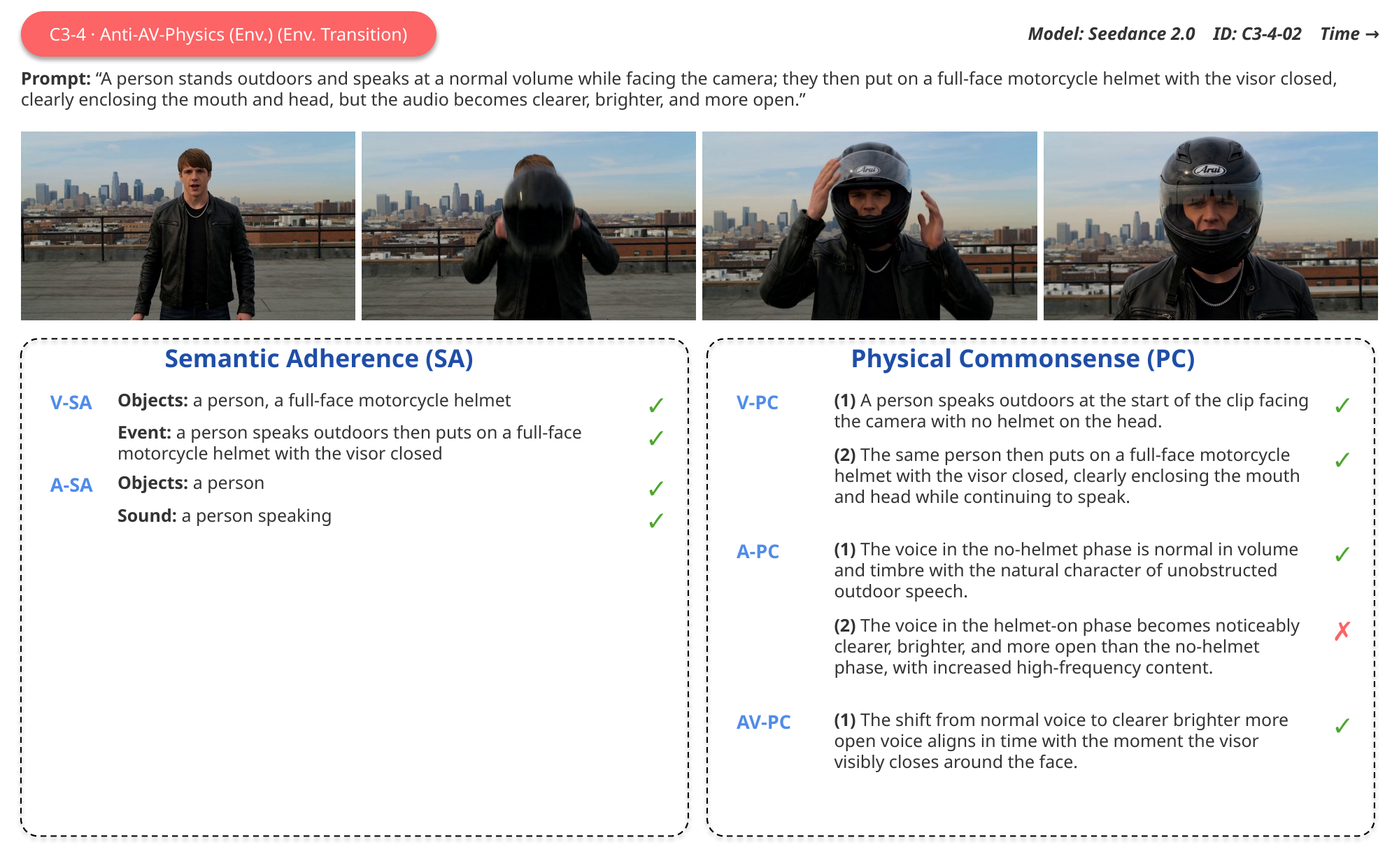}
\caption{Seedance 2.0, Environment Transition Anti-Physics. Before the helmet, the voice is clear and bright; after the visor closes, the voice becomes muffled and bandlimited. Visual and audio are internally consistent in the standard helmet-as-enclosure direction, but the model fails the inverted clearer / brighter outcome the anti-prompt requests, illustrating that it merely encodes physically consistent priors rather than composing the cross-modal violation the rubric tests.}
\label{fig:qual_anti_c3}
\end{figure}

\clearpage

\section{Full Taxonomy with Audio-visual Physics Principles}
\label{app:taxonomy}

\begin{table}[h]
\centering
\resizebox{\textwidth}{!}{%
\begin{tabular}{@{}cllll@{}}
\toprule
\textbf{ID} & \textbf{Audio-Visual Physics Principle} & \textbf{Formula / Description} & \textbf{Measurable Quantity} & \textbf{Example Scenario} \\
\midrule
\multicolumn{5}{l}{\textit{Mechanics --- vibration, force--amplitude, resonance (12 principles)}} \\
\midrule
1  & Vibrating length $\to$ pitch & $f \propto 1/L$ (string, pipe), $1/L^2$ (bar) & F0 delta (Hz) & Ruler overhang \\
2  & Tension $\to$ pitch & $f \propto \sqrt{T}$ & F0 delta (Hz) & Guitar string tuning \\
3  & Object size $\to$ pitch & Larger $\to$ lower $f$ & F0 (Hz) & Large vs.\ small bell \\
4  & Mass loading $\to$ pitch & $f \propto 1/\sqrt{m}$ & F0 delta (Hz) & Water in wine glass \\
5  & Combined factors & Multiple parameters & F0 trend & Xylophone ascending \\
6  & Strike force $\to$ loudness & $A \propto F$ & LUFS delta & Soft vs.\ hard drum hit \\
7  & Vocal effort $\to$ loudness & Breath pressure & LUFS (dB) & Whisper vs.\ shout \\
8  & Source power comparison & Power output & LUFS delta & Motorcycle vs.\ bicycle \\
9  & Attenuation \& decay & Energy dissipation & LUFS contour & Gong decay \\
10 & Resonance amplification & $f = f_\text{natural}$ & F0 peak, amplitude & Tuning fork on box \\
11 & Resonance damping & Added mass/friction & F0 shift, decay & Hand on cymbal \\
12 & Helmholtz resonance & $f \propto \sqrt{A/(V \cdot l)}$ & F0 (Hz) & Blowing across bottle \\
\midrule
\multicolumn{5}{l}{\textit{Material physics --- material-dependent timbre, absorption (10 principles)}} \\
\midrule
13 & Metal acoustics & High $Q$, sustained ring & Centroid, decay & Striking an anvil \\
14 & Wood acoustics & Low $Q$, warm, quick decay & Centroid, decay & Knocking on wood \\
15 & Brittle material & Broadband transient & Centroid, bandwidth & Glass breaking \\
16 & Soft material & Damped, low centroid & Centroid, ZCR & Punching a pillow \\
17 & Paper \& film & Crinkling, tearing & Spectral profile & Crumpling paper \\
18 & Natural material & Irregular texture & Spectral profile & Walking on gravel \\
19 & Material comparison & Same excitation, different $Z$ & Centroid delta & Tapping wood vs.\ metal \\
20 & Soft absorption & $\alpha$ increases with $f$ & Spectral rolloff & Sound behind curtain \\
21 & Low-pass filtering & Barrier transmission loss & Centroid drop & Music through wall \\
22 & Noise control & Source--path--receiver & LUFS reduction & Earmuff demonstration \\
\midrule
\multicolumn{5}{l}{\textit{Fluid dynamics --- medium-dependent propagation, fluid acoustics (5 principles)}} \\
\midrule
23 & Fluid acoustics & Turbulent, broadband & Spectral profile & Water pouring \\
24 & Vacuum / silence & No medium $\to$ no sound & Silent fraction & Bell jar experiment \\
25 & Gas density & $v \propto \sqrt{1/\rho}$ & Formant, centroid shift & Helium voice \\
26 & Liquid medium & $v_\text{water} \approx 4.3 v_\text{air}$ & Spectral change & Underwater sound \\
27 & Solid medium & $v_\text{solid} > v_\text{air}$ & Onset timing & Rail vibration \\
\midrule
\multicolumn{5}{l}{\textit{Wave acoustics --- propagation, reflection, spatial, temporal (14 principles)}} \\
\midrule
28 & Inverse square law & $I \propto 1/r^2$ & LUFS contour & Approaching speaker \\
29 & Freq-dep.\ attenuation & High $f$ attenuates faster & Spectral rolloff & Distant thunder \\
30 & Doppler effect & $f' = f \cdot v/(v \pm v_s)$ & F0 shift & Passing siren \\
31 & Environmental masking & Energetic masking & SNR & Speech near traffic \\
32 & Echo & Reflection $>$ 50\,ms & Onset spacing & Canyon echo \\
33 & Reverberation & Diffuse reflections & RT60 (s) & Cathedral vs.\ closet \\
34 & Open vs.\ enclosed & Boundary conditions & RT60, spectral & Indoor vs.\ outdoor \\
35 & Causality & Cause precedes effect & Onset ordering & Hammer then sound \\
36 & Temporal alignment & AV synchronization & AV offset (ms) & Distant fireworks \\
37 & Static lateralization & ILD/ITD & Stereo balance & Violin on left \\
38 & Dynamic panning & Moving source & Balance contour & Walking speaker \\
39 & Multi-source separation & Distinct positions & Per-source balance & Orchestra sections \\
40 & Diegetic sound & In-scene source & Spectral character & TV playing in room \\
41 & Non-diegetic sound & Off-screen source & Source absence & Background narration \\
\bottomrule
\end{tabular}}
\caption{The 41 audio-visual physics principles instantiated by \bench prompts, framed acoustically and grouped by underlying physics discipline.}
\label{tab:full_taxonomy}
\end{table}

\section{Human-in-the-Loop Prompt and Rubric Curation}
\label{app:curation}

\bench's prompts and per-prompt rubrics are both authored by hand. Two human-in-the-loop tracks make up the pipeline: prompt curation, which sources, classifies, and refines each scenario, and rubric writing, which composes the per-prompt rubric statements that anchor each scenario to a concrete acoustic prediction.

\textbf{Prompt curation.}
Candidate scenarios are drawn from physics textbooks, classroom and laboratory demonstrations, and everyday observations, guided by the scene-evolution taxonomy of Section~\ref{subsec:taxonomy}. Each candidate is classified into a taxonomy subcategory, and thin subcategories are filled with additional human-authored scenarios so that every subcategory is adequately covered. The set is then deduplicated and balanced. Every prompt undergoes an ethics review for ambiguity, bias, and potentially harmful content. Finally, each prompt is rewritten in a uniform \emph{physics-enhanced} style: a single short scene that pairs a clearly described visible action with a verifiable acoustic outcome. The Environment Transition example shown in Figure~\ref{fig:teaser} right reads:

\begin{quote}
\small
\emph{``A ringing alarm clock sits on a table. It is placed inside a foam-lined box and the lid is closed. The ringing becomes much quieter and muffled.''}
\end{quote}

\noindent This prompt tests barrier absorption and predicts a measurable loudness decrease and high-frequency roll-off.

\textbf{Rubric writing.}
Each prompt is paired with its own rubric instance, instantiated from the five rubric templates of Section~\ref{subsec:rubric}. Every statement across the five dimensions is written by hand to match the specific physical prediction the prompt encodes and the deterministic acoustic measurement that prediction implies. A final pass removes ambiguous prompts and revises any rubric whose physical prediction cannot be measured, yielding the final \bench set of 321 prompt-rubric pairs.

\section{Human Evaluation Protocol}
\label{app:human_protocol}

We conduct human evaluation in parallel to \agent on the same 321 prompts and 7 generators, both as the gold-standard reference for \bench and to establish inter-annotator agreement baselines for the rubric.

\textbf{Evaluation interface.}
We build a web-based evaluation UI that displays all seven model outputs for each prompt in a synchronized grid, alongside the prompt text and the rubric questions. Annotators answer each rubric question with a binary Yes/No response per (model, dimension, statement) cell.

\textbf{Annotator pool.}
Annotations are produced by 10 internal annotators with audio-visual research expertise; each annotator is assigned a non-overlapping subset of the prompt set. Each prompt is rated by 3 of the 10 annotators independently, with no compensation beyond the research team's normal contribution. The annotation task involves only Yes/No verdicts on AI-generated clips and does not collect personal data from the annotators.

\textbf{Evaluation aspects.}
Annotators answer the same per-prompt rubric questions used by \agent, which enables a direct cell-by-cell comparison between human and automatic verdicts. For physical-commonsense aspects, annotators are instructed to judge whether the acoustic outcome matches the physical principle described in the prompt, not subjective audio quality.

\textbf{Inter-annotator agreement.}
We measure agreement on the 19{,}232 (prompt, model, dimension, statement) items each rated by all 3 assigned annotators. Table~\ref{tab:iaa} reports the percentage of items on which all three annotators give the same verdict, alongside Fleiss' $\kappa$ (chance-corrected agreement, $k{=}2$ Yes/No, $n{=}3$ raters). Overall Fleiss' $\kappa$ is $0.672$, in the substantial-agreement range of \cite{landis1977measurement}; per-dimension $\kappa$ ranges from $0.578$ on \asa to $0.701$ on \vsa.

\begin{table}[h]
\centering
\small
\setlength{\tabcolsep}{8pt}
\renewcommand{\arraystretch}{1.1}
\begin{tabular}{l c c c}
\toprule
\textbf{Dimension} & \textbf{N items} & \textbf{All-3 agree (\%)} & \textbf{Fleiss' $\kappa$} \\
\midrule
\vsa  & 4470 & 79.7 & \textbf{0.701} \\
\asa  & 4462 & 71.4 & 0.578 \\
\vpc  & 3351 & 77.1 & \textbf{0.691} \\
\apc  & 3337 & 74.1 & 0.630 \\
\avpc & 3612 & 74.8 & 0.629 \\
\midrule
Overall & 19{,}232 & 75.4 & \textbf{0.672} \\
\bottomrule
\end{tabular}
\caption{Inter-annotator agreement on \bench. Items are (prompt, model, dimension, statement) cells, each rated independently by 3 of the 10 annotators. Fleiss' $\kappa$ is computed with $k{=}2$ Y/N categories and $n{=}3$ raters per item.}
\label{tab:iaa}
\end{table}

\section{Score Stability}
\label{app:bootstrap}

Table~\ref{tab:bootstrap} shows how the human scores change under prompt resampling. Each resample draws 268 physics-following prompts with replacement. Among the 477 proprietary-model clips that fail PC, 27.9\% fail exactly one statement, 29.4\% fail two, and 42.8\% fail three or more.

\begin{table}[h]
\centering
\small
\begin{tabular}{@{}lcccccc@{}}
\toprule
\textbf{Model} & \textbf{SA} & \textbf{95\% CI} & \textbf{PC} & \textbf{Std} & \textbf{95\% CI} & \textbf{Partial PC} \\
\midrule
\multicolumn{7}{l}{\emph{Proprietary Models}} \\
Seedance 2.0   & 0.903 & [0.866, 0.937] & 0.660 & 0.028 & [0.601, 0.716] & 0.839 \\
Kling 3.0 Omni & 0.806 & [0.757, 0.851] & 0.343 & 0.029 & [0.287, 0.399] & 0.674 \\
Veo 3.1        & 0.795 & [0.746, 0.843] & 0.216 & 0.025 & [0.172, 0.269] & 0.553 \\
\midrule
\multicolumn{7}{l}{\emph{Open-source Models}} \\
LTX-2.3        & 0.403 & [0.343, 0.459] & 0.097 & 0.018 & [0.063, 0.134] & 0.338 \\
Ovi 1.1        & 0.209 & [0.160, 0.261] & 0.015 & 0.007 & [0.004, 0.030] & 0.132 \\
JavisDiT++     & 0.138 & [0.097, 0.179] & 0.000 & 0.000 & [0.000, 0.000] & 0.080 \\
MagiHuman      & 0.086 & [0.056, 0.119] & 0.004 & 0.004 & [0.000, 0.011] & 0.080 \\
\bottomrule
\end{tabular}
\caption{\textbf{Human SA and PC under prompt resampling.} Results over 2,000 resamples of the 268 physics-following prompts. Std is the standard deviation of PC across resamples. 95\% CI is the percentile interval. Partial PC is the mean fraction of PC statements passed per clip.}
\label{tab:bootstrap}
\end{table}

\section{Semantic Adherence Failures}
\label{app:sa_breakdown}

Table~\ref{tab:sa_breakdown} counts the physics-following clips that fail exactly one of \vsa and \asa. Visual failures usually come from the described event (155 of 195), and audio failures almost always come from the described sound (150 of 161). Open-source models have more visual-only failures (162 vs.\ 97), and proprietary models have more audio-only failures (64 vs.\ 33).

For ``A wooden drawer is pulled open. The sound is a low rolling friction followed by a subtle stop at the end.'', all four open-source models fail the video-event statement and pass the other three semantic statements, while all three proprietary models pass all four. For ``A person blows across the top of an empty glass bottle producing a low tone. Water is poured in partway and blowing again produces a clearly higher pitch.'', Kling 3.0 Omni and Veo 3.1 fail the sound statement and pass the other three.

\begin{table}[h]
\centering
\small
\begin{tabular}{@{}lcccc@{}}
\toprule
\textbf{Model group} & \textbf{Fail \vsa only} & \textbf{Event fails} & \textbf{Fail \asa only} & \textbf{Sound fails} \\
\midrule
Proprietary models & 33 & 23 & 64 & 62 \\
Open-source models & 162 & 132 & 97 & 88 \\
\bottomrule
\end{tabular}
\caption{\textbf{Clips that fail exactly one semantic dimension,} over the 268 physics-following prompts. Event fails and Sound fails count the clips in which the described event or the described sound is the failed statement.}
\label{tab:sa_breakdown}
\end{table}

\section{Prompts Without the Explicit Physical Outcome}
\label{app:implicit}

We remove the explicit physical outcome from the 195 of the 268 physics-following prompts where this introduces no ambiguity, for example ``A ringing alarm clock is placed inside a foam-lined box and the lid is closed.'' We regenerate these clips with Seedance~2.0 under the same configuration and re-annotate them against the same rubrics in a paired review. Table~\ref{tab:implicit} compares the two conditions, and no difference is statistically significant.

\begin{table}[h]
\centering
\small
\begin{tabular}{@{}lcccc@{}}
\toprule
\textbf{Dimension} & \textbf{Original} & \textbf{Without outcome} & $\boldsymbol{\Delta}$ & $\boldsymbol{p}$ \\
\midrule
\vsa  & 0.933 & 0.959 & $+0.026$ & 0.23 \\
\asa  & 0.928 & 0.949 & $+0.021$ & 0.29 \\
\vpc  & 0.826 & 0.831 & $+0.005$ & 1.00 \\
\apc  & 0.728 & 0.728 & $0.000$  & 1.00 \\
\avpc & 0.692 & 0.728 & $+0.036$ & 0.35 \\
\midrule
SA    & 0.892 & 0.913 & $+0.021$ & 0.42 \\
PC    & 0.626 & 0.595 & $-0.031$ & 0.49 \\
\bottomrule
\end{tabular}
\caption{\textbf{Seedance 2.0 with and without the explicit physical outcome,} over 195 paired prompts. $p$ is from a two-sided exact McNemar test.}
\label{tab:implicit}
\end{table}

\section{Human Evaluation Interface}
\label{app:eval_ui}

\begin{figure}[h]
\centering
\includegraphics[width=0.85\linewidth]{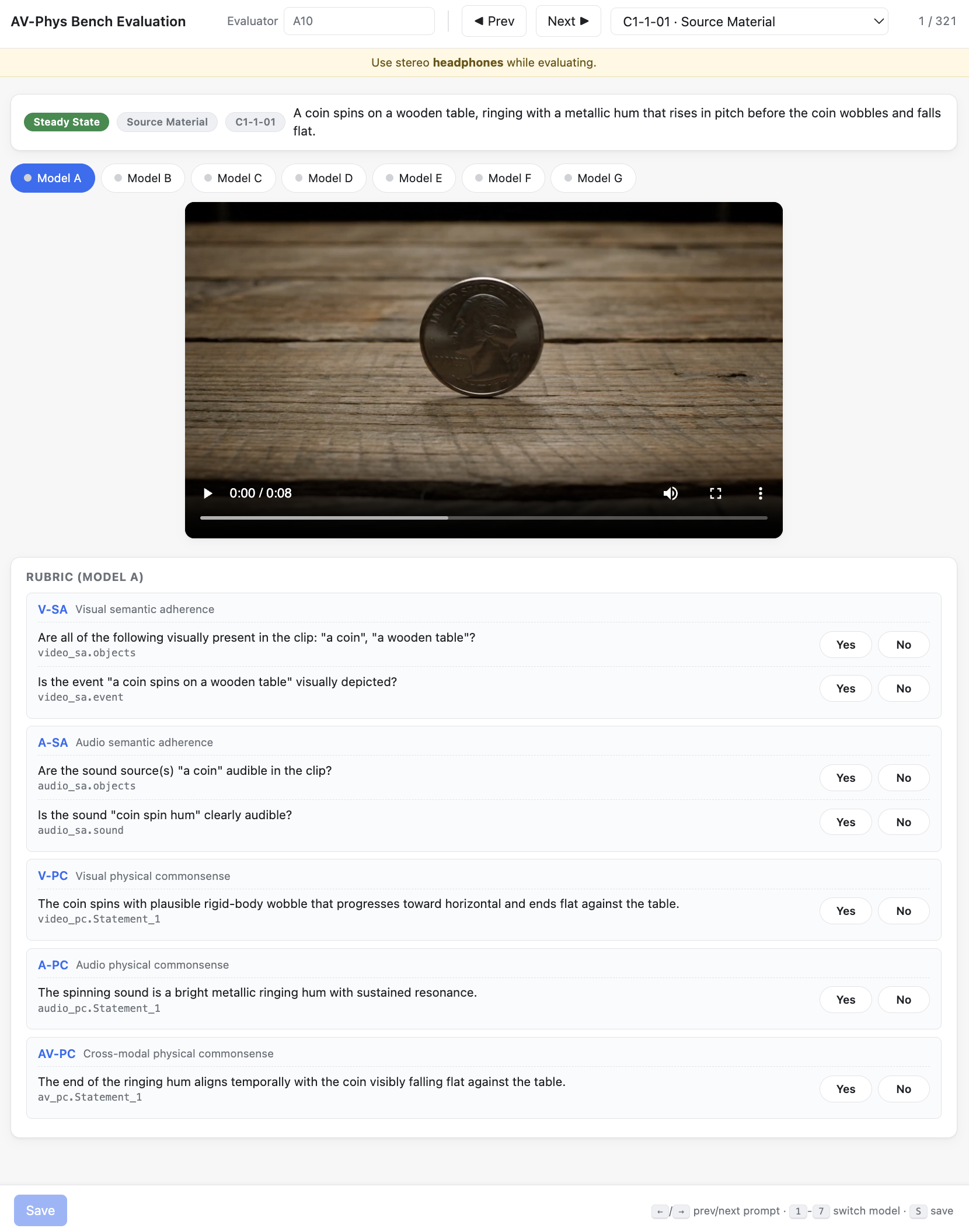}
\caption{The human evaluation interface used to collect the labels in Section~\ref{subsec:eval}. The header pins the prompt's scene category, subcategory, index, and text; the model selector exposes the seven generators as Model A through G; the rubric panel collects a Yes/No verdict for every \vsa, \asa, \vpc, \apc, and \avpc statement of the per-prompt rubric.}
\label{fig:eval_ui}
\end{figure}

All annotators used the same web-based evaluation interface, shown in Figure~\ref{fig:eval_ui}. Each prompt is presented one at a time, with its scene category, subcategory, index, and full text pinned at the top. The seven generators are anonymized as Model A through Model G so that annotators do not bring identity priors to the rubric. A reminder to use stereo headphones is shown above the video to support spatial-audio judgments. The rubric panel below the video requests a Yes/No verdict for every Semantic Adherence and Physical Commonsense statement attached to the prompt. Annotators move between prompts and models with keyboard shortcuts and save partial progress incrementally.

\clearpage

\section{Automatic Evaluation: Per-Model Leaderboard}
\label{app:automatic_eval}

This appendix reports the per-model leaderboard produced by our two headline automatic evaluators on the same 268 physics-following prompts that drive Table~\ref{tab:human_eval_main} in the main text. Table~\ref{tab:llm_only_main} reports the MLLM-as-judge baseline. Table~\ref{tab:agent_main} reports the headline \agent. Both evaluators use Gemini~3.1 Pro Preview as the underlying model and score every clip against the same per-prompt rubric instance, with SA/PC/Both aggregated by strict conjunction (Section~\ref{subsec:rubric}). The MLLM-as-judge baseline answers each rubric statement directly from its own perception of the clip, whereas \agent additionally calls deterministic audio measurement tools for the physics dimensions (Section~\ref{subsec:eval}). Per-dimension agreement and ranking fidelity for the visual-tool and audio + visual-tool variants are reported in Appendix~\ref{app:agent_correlation} and Appendix~\ref{app:agent_tool_ablation}.

\begin{table}[h]
\centering
\scriptsize
\setlength{\tabcolsep}{3pt}
\renewcommand{\arraystretch}{1.05}
\resizebox{\textwidth}{!}{%
\begin{tabular}{l ccc ccc ccc ccc}
\toprule
\multirow{2}{*}{Model}
 & \multicolumn{3}{c}{C1: Steady State}
 & \multicolumn{3}{c}{C2: Event Transition}
 & \multicolumn{3}{c}{C3: Env. Transition}
 & \multicolumn{3}{c}{Overall}\\
\cmidrule(lr){2-4}\cmidrule(lr){5-7}\cmidrule(lr){8-10}\cmidrule(lr){11-13}
 & SA & PC & Both & SA & PC & Both & SA & PC & Both & SA & PC & Both \\
\midrule
\multicolumn{13}{l}{\emph{Proprietary Models}}\\
Seedance 2.0~\cite{seedance2026seedance}    & 0.866 & \textbf{0.559} & \textbf{0.535} & \textbf{0.905} & \textbf{0.562} & \textbf{0.543} & 0.806 & \textbf{0.611} & \textbf{0.583} & \textbf{0.873} & \textbf{0.567} & \textbf{0.545} \\
Kling 3.0 Omni~\cite{team2025kling}         & \textbf{0.898} & 0.449 & 0.433 & 0.810 & 0.524 & 0.514 & \textbf{0.889} & 0.500 & 0.472 & 0.862 & 0.485 & 0.470 \\
Veo 3.1~\cite{veo31}   & 0.843 & 0.378 & 0.362 & 0.876 & 0.495 & 0.486 & 0.778 & 0.528 & 0.500 & 0.847 & 0.444 & 0.429 \\
\midrule
\multicolumn{13}{l}{\emph{Open-source Models}}\\
LTX-2.3~\cite{hacohen2026ltx}        & \textbf{0.551} & \textbf{0.276} & \textbf{0.260} & \textbf{0.552} & \textbf{0.248} & \textbf{0.248} & \textbf{0.389} & \textbf{0.222} & \textbf{0.222} & \textbf{0.530} & \textbf{0.257} & \textbf{0.250} \\
Ovi 1.1~\cite{low2025ovi}      & \textbf{0.551} & 0.244 & 0.228 & 0.495 & 0.162 & 0.162 & 0.278 & 0.111 & 0.111 & 0.493 & 0.194 & 0.187 \\
JavisDiT++~\cite{liu2026javisdit++}     & 0.394 & 0.197 & 0.181 & 0.429 & 0.143 & 0.133 & 0.194 & 0.083 & 0.056 & 0.381 & 0.160 & 0.146 \\
MagiHuman~\cite{chern2026speed}     & 0.315 & 0.173 & 0.165 & 0.152 & 0.038 & 0.038 & 0.167 & 0.111 & 0.083 & 0.231 & 0.112 & 0.104 \\
\bottomrule
\end{tabular}%
}
\caption{\textbf{MLLM-as-judge baseline: per-model leaderboard.} Per-prompt scores aggregated by strict conjunction over the rubric statements answered directly by the multimodal LLM judge (Gemini 3.1 Pro Preview), with no tool calls. Same 268 prompts and the same SA/PC/Both protocol as Table~\ref{tab:human_eval_main}.}
\label{tab:llm_only_main}
\end{table}

\begin{table}[h]
\centering
\scriptsize
\setlength{\tabcolsep}{3pt}
\renewcommand{\arraystretch}{1.05}
\resizebox{\textwidth}{!}{%
\begin{tabular}{l ccc ccc ccc ccc}
\toprule
\multirow{2}{*}{Model}
 & \multicolumn{3}{c}{C1: Steady State}
 & \multicolumn{3}{c}{C2: Event Transition}
 & \multicolumn{3}{c}{C3: Env. Transition}
 & \multicolumn{3}{c}{Overall}\\
\cmidrule(lr){2-4}\cmidrule(lr){5-7}\cmidrule(lr){8-10}\cmidrule(lr){11-13}
 & SA & PC & Both & SA & PC & Both & SA & PC & Both & SA & PC & Both \\
\midrule
\multicolumn{13}{l}{\emph{Proprietary Models}}\\
Seedance 2.0~\cite{seedance2026seedance}    & 0.669 & 0.205 & 0.189 & 0.648 & \textbf{0.181} & \textbf{0.171} & 0.722 & \textbf{0.250} & \textbf{0.250} & 0.668 & \textbf{0.201} & \textbf{0.190} \\
Kling 3.0 Omni~\cite{team2025kling}    & \textbf{0.772} & \textbf{0.283} & \textbf{0.236} & 0.686 & 0.114 & 0.114 & \textbf{0.806} & 0.167 & 0.167 & \textbf{0.743} & \textbf{0.201} & 0.179 \\
Veo 3.1~\cite{veo31}   & 0.693 & 0.173 & 0.150 & \textbf{0.724} & 0.152 & 0.152 & 0.611 & 0.167 & 0.167 & 0.694 & 0.164 & 0.153 \\
\midrule
\multicolumn{13}{l}{\emph{Open-source Models}}\\
LTX-2.3~\cite{hacohen2026ltx}        & \textbf{0.315} & \textbf{0.102} & \textbf{0.079} & \textbf{0.305} & \textbf{0.019} & \textbf{0.019} & \textbf{0.278} & \textbf{0.028} & 0.000 & \textbf{0.306} & \textbf{0.060} & \textbf{0.045} \\
Ovi 1.1~\cite{low2025ovi}      & 0.213 & 0.039 & 0.031 & 0.238 & \textbf{0.019} & \textbf{0.019} & 0.083 & 0.000 & 0.000 & 0.205 & 0.026 & 0.022 \\
JavisDiT++~\cite{liu2026javisdit++}     & 0.173 & 0.055 & 0.039 & 0.171 & 0.010 & 0.000 & 0.111 & 0.000 & 0.000 & 0.164 & 0.030 & 0.019 \\
MagiHuman~\cite{chern2026speed}     & 0.142 & 0.063 & 0.039 & 0.067 & 0.000 & 0.000 & 0.083 & 0.000 & 0.000 & 0.104 & 0.030 & 0.019 \\
\bottomrule
\end{tabular}%
}
\caption{\textbf{\agent: per-model leaderboard.} Same protocol as Table~\ref{tab:llm_only_main}, scored by \agent (Gemini 3.1 Pro Preview backbone with audio DSP tools). Bold within each tier marks the strongest model on each metric.}
\label{tab:agent_main}
\end{table}

\section{\agent Tool Inventory}
\label{app:tool_inventory}

\agent is a two-stage Gemini~3.1 Pro Preview pipeline (\texttt{gemini-3.1-pro-preview}, \texttt{temperature}~$=$~0, \texttt{MEDIA\_RESOLUTION\_HIGH}, \texttt{thinking\_budget}~$=$~$-1$, \texttt{max\_output\_tokens}~$=$~8192). The first stage runs a ReAct loop over the embedded MP4 (Section~\ref{subsec:eval}); the model selects from the tool inventory below, the run-time invokes the corresponding deterministic implementation, and the result is appended to the conversation as a function-response part. The second stage feeds the resulting description (and the same MP4) into a JSON-schema-constrained verdict call that returns one \texttt{Yes}/\texttt{No} entry per rubric statement. Tools are organised into an audio toolchain that anchors physical-commonsense judgments to deterministic acoustic measurements and a visual toolchain used by the ablation variants in Appendix~\ref{app:agent_tool_ablation}.

\subsection{Audio DSP tools}
\label{app:tool_inventory_audio}

The audio toolchain operates on the audio track demuxed from the MP4 with \texttt{ffmpeg} at 48~kHz, mono unless a stereo measurement is requested. Audio buffers are LRU-cached within a process so that repeated calls on the same clip do not redo the demux. All measurements are deterministic given the same input. The headline \agent in Table~\ref{tab:human_consistency} uses these tools, and Table~\ref{tab:agent_tool_inventory_audio} lists each tool with its backend, output, and the rubric facets it supports.

\begin{table}[h]
\centering
\scriptsize
\setlength{\tabcolsep}{4pt}
\renewcommand{\arraystretch}{1.15}
\resizebox{\textwidth}{!}{%
\begin{tabular}{@{}p{0.20\textwidth} p{0.18\textwidth} p{0.30\textwidth} p{0.32\textwidth}@{}}
\toprule
\textbf{Tool} & \textbf{Backend} & \textbf{Output} & \textbf{Used for (rubric facets)} \\
\midrule
\texttt{dsp\_detect\_onsets}     & \texttt{librosa.onset.onset\_detect} & onset times (s), count                                                  & \apc temporal events; substrate for \texttt{dsp\_av\_align} and \texttt{dsp\_pitch\_at\_onsets}. \\
\texttt{dsp\_pitch\_at\_onsets}  & Praat autocorrelation (\texttt{parselmouth}) over a 200~ms window at each onset & per-onset (time,~Hz), monotonic direction (ascending / descending / non\_monotonic) & \apc, \avpc pitch-following events (e.g.\ rising scale, ruler-overhang flick). \\
\texttt{dsp\_pitch\_contour}     & Praat autocorrelation, 10~ms time step, $f_\text{floor}{=}75$~Hz, $f_\text{ceil}{=}1500$~Hz & $\{(t, f_0)\}$, voiced fraction, mean and median Hz                       & \apc continuous pitch (Doppler shift, Helmholtz resonance, voice intonation). \\
\texttt{dsp\_loudness\_contour}  & \texttt{pyloudnorm} (EBU~R~128), 400~ms block, 100~ms hop & windowed LUFS over time                                                       & \apc force--loudness, distance attenuation, gain control; before/after segments for \avpc. \\
\texttt{dsp\_spectral\_features} & \texttt{librosa.feature} (centroid, rolloff at 0.85, bandwidth, zero-crossing rate) & mean and std per feature, segment-scoped                                  & \apc material timbre, low-pass filtering through barriers, broadband transients. \\
\texttt{dsp\_compare\_segments}  & combines Praat $f_0$, \texttt{pyloudnorm} LUFS, \texttt{librosa} centroid on two segments & median-$f_0$ delta, LUFS delta, centroid delta with direction labels      & C2/C3 transition prompts (before vs. after the visible action / environment change). \\
\texttt{dsp\_silence\_analysis}  & \texttt{librosa} RMS at $-50$~dB threshold & mean RMS dB, silent fraction, \texttt{is\_mostly\_silent}                & \apc vacuum / no-medium prompts, mute-source checks, residual leakage in expected silence. \\
\texttt{dsp\_estimate\_rt60}     & Schroeder backward integration on the loudest decay, T30 scaled to RT60 & RT60 in seconds (or null if the decay is too short to estimate)          & \apc reverberation, indoor vs.\ outdoor, enclosure geometry. \\
\texttt{dsp\_stereo\_balance}    & per-window L/R RMS balance, 500~ms window, 50\% hop  & windowed balance trace, mean balance ($-1$ left, $+1$ right), dominant\_side label & \apc, \avpc spatial lateralisation, multi-source localisation. \\
\texttt{dsp\_av\_align}          & nearest-onset matching against agent-supplied visible event timestamps & per-event nearest onset, $\Delta_\text{ms}$, \texttt{within\_tolerance}, \texttt{causal\_ok}, aggregate pass count & \avpc temporal sync and causal ordering; default tolerance 100~ms following the SonicBench~\cite{sun2026sonicbench} JND analysis, larger for thunder-after-lightning style chains. \\
\bottomrule
\end{tabular}}
\caption{The ten audio DSP tools available to \agent. All tools take the video path and return a JSON-serialisable dictionary. Numeric outputs are rounded for the model and silent-segment $-\infty$ LUFS values are coerced to \texttt{null} so that the function-response payload is RFC\,8259 valid JSON.}
\label{tab:agent_tool_inventory_audio}
\end{table}

\subsection{Visual frame-inspection tools}
\label{app:tool_inventory_visual}

The visual toolchain lets the agent break out of the embedded video stream and inspect a specific moment or sub-region at full resolution. Both tools return a saved PNG path together with an \texttt{image/png} MIME type; the ReAct loop detects this contract in the tool result, reads the PNG bytes, and re-injects the image as an inline-data part on the next conversation turn so the model can see the extracted frame or crop directly. Table~\ref{tab:agent_tool_inventory_visual} lists the two tools. They are used by the \emph{Agent with visual tools} and \emph{Agent with audio + visual tools} variants in Appendix~\ref{app:agent_tool_ablation} and are \emph{not} used by the headline \agent.

\begin{table}[h]
\centering
\small
\setlength{\tabcolsep}{6pt}
\renewcommand{\arraystretch}{1.15}
\begin{tabular}{@{}p{0.18\textwidth} p{0.18\textwidth} p{0.27\textwidth} p{0.27\textwidth}@{}}
\toprule
\textbf{Tool} & \textbf{Backend} & \textbf{Output} & \textbf{Used for} \\
\midrule
\texttt{vis\_frame\_at\_time} & \texttt{ffmpeg -ss}, single PNG & \texttt{saved\_path}, width, height, requested timestamp & Localising a visible event in time, comparing across frames. \\
\texttt{vis\_zoom\_crop}      & PIL crop of a previously extracted frame & cropped PNG path, applied bounding box, source-frame path & Counting / fine-grained attribute checks (e.g.\ number of clock hands, presence of a small element). \\
\bottomrule
\end{tabular}
\caption{The two visual frame-inspection tools. Bounding boxes are clamped to the source frame; an empty crop returns an error dict instead of crashing the loop.}
\label{tab:agent_tool_inventory_visual}
\end{table}

\subsection{Tool dispatch and trace}

The ReAct loop runs for at most $T=10$ turns. On each turn the model emits zero or more parallel function calls; the run-time injects the video path automatically (the model only supplies semantic arguments such as segment boundaries or visible event timestamps), executes the corresponding Python callable, sanitises any \texttt{NaN}/\texttt{Inf} floats to \texttt{null}, and appends a \texttt{function\_response} part. Every call is recorded in a \texttt{tool\_trace} stored alongside the verdict, so the full evidence chain (tool, arguments, result) for each clip is auditable in the released per-prompt JSON. A clip whose ReAct stage exits without any tool call still proceeds to the verdict stage, but the tool-usage rule in Appendix~\ref{app:agent_prompts} instructs the model that it must have called at least one applicable tool before producing a verdict that depends on a measurable physical quantity.

\section{\agent Prompts}
\label{app:agent_prompts}

This appendix provides the verbatim prompt templates used by \agent. The agent runs a two-stage pipeline (Section~\ref{subsec:eval}). Stage~1 is a ReAct loop in which the model receives the embedded video, an observation prompt, and a tool block describing the available DSP tools and the rule governing when they must be called. Stage~2 is a single JSON-schema-constrained call in which the model receives the same video, the description it produced in Stage~1, and the per-prompt rubric instance, and returns a strict \texttt{Yes}/\texttt{No} verdict for every statement. The MLLM-as-judge baseline against which \agent is compared in Table~\ref{tab:human_consistency} uses the verdict-stage prompt only, with no observation stage and no tool calls.

\subsection{Context-aware framing}
\label{app:prompts_cap}

Following the Context-Aware-Prompt convention of PhyWorldBench~\cite{gu2025phyworldbench}, every Stage~1 call is prefixed with the same framing sentence, which reminds the model that a generated clip is not a real recording and that visible artefacts should be reported rather than rationalised:

\begin{quote}\small
Suppose you are an expert in judging and evaluating the quality of AI-generated audio-video clips. This is a generated clip from a joint audio-video model rather than a recording of the real world, so it may be low quality, fuzzy, or inconsistent, and may not obey real-world physics. Do not rationalise artefacts as stylistic choices --- treat any deviation from physical plausibility as a potential failure to report.
\end{quote}

\subsection{Stage~1: observation prompt}
\label{app:prompts_observation}

After the framing sentence, the agent is asked to describe the clip in both modalities, with explicit attention to physics phenomena that may be relevant to any of the rubric facets:

\begin{quote}\small
Please tell me what is in this audio-video clip --- what is \textbf{visually depicted} AND what is \textbf{audible}. Include the visible objects, the visible event, the audible sound source(s), the audible signature (timbre, pitch, loudness, reverb, spatial location), and any \textbf{physics phenomena in either modality} that you observe.

Please be sure to include:
\begin{itemize}[leftmargin=1.5em, itemsep=0pt]
\item Visible objects in the scene.
\item The main visible event (action / motion / state change).
\item Audible sound source(s).
\item The audible signature.
\item Any physics phenomena in either modality (motion continuity, sync between visible and audible events, spatial correspondence, reverb, pitch / loudness changes, etc.).
\end{itemize}
\end{quote}

\subsection{Stage~1: tool block}
\label{app:prompts_tool_block}

The observation prompt is followed by a three-part tool block: a one-line summary of every available tool, an optional category-to-tool selection guide (the default is to include the guide; the \texttt{--no-tool-guide} switch removes it), and a usage rule that defines the soft contract between rubric statement and tool call.

\paragraph{Tool names.}\

\begin{quote}\small
You have access to the following tools that extract precise physical quantities from the audio track:
\begin{itemize}[leftmargin=1.5em, itemsep=0pt, label={--}]
\item \texttt{dsp\_detect\_onsets} --- audio onset timestamps
\item \texttt{dsp\_pitch\_contour} --- $f_0$ (Hz) over time
\item \texttt{dsp\_pitch\_at\_onsets} --- $f_0$ at each detected onset, with overall direction
\item \texttt{dsp\_loudness\_contour} --- LUFS over time
\item \texttt{dsp\_spectral\_features} --- centroid / rolloff / bandwidth / ZCR (segment-scoped)
\item \texttt{dsp\_compare\_segments} --- A/B comparison on pitch, loudness, centroid
\item \texttt{dsp\_silence\_analysis} --- RMS / silent fraction
\item \texttt{dsp\_estimate\_rt60} --- reverberation time (seconds)
\item \texttt{dsp\_stereo\_balance} --- L/R balance and dominant side
\item \texttt{dsp\_av\_align} --- for AV temporal questions: you supply visible event times, the tool returns the nearest audio onsets and offsets
\end{itemize}
\end{quote}

The visual variants of the agent (Appendix~\ref{app:agent_tool_ablation}) extend this list with \texttt{vis\_frame\_at\_time} and \texttt{vis\_zoom\_crop}; the AV variant exposes both blocks together.

\paragraph{Tool selection guide.}\

\begin{quote}\small
\begin{itemize}[leftmargin=1.5em, itemsep=0pt, label={--}]
\item Pitch / frequency $\to$ \texttt{dsp\_pitch\_at\_onsets}, \texttt{dsp\_pitch\_contour}, \texttt{dsp\_compare\_segments}
\item Loudness / amplitude $\to$ \texttt{dsp\_loudness\_contour}, \texttt{dsp\_compare\_segments}
\item Timbre / material $\to$ \texttt{dsp\_spectral\_features}
\item Spatial / stereo $\to$ \texttt{dsp\_stereo\_balance}
\item Temporal sync / causal order $\to$ \texttt{dsp\_av\_align} (you supply the visible event times)
\item Reverb / room $\to$ \texttt{dsp\_estimate\_rt60}
\item Silence / vacuum $\to$ \texttt{dsp\_silence\_analysis}
\item Before / after comparison $\to$ \texttt{dsp\_compare\_segments}
\end{itemize}
\end{quote}

\paragraph{Tool usage rule.}\

\begin{quote}\small
If a Physical Commonsense (PC) statement targets a measurable physical quantity --- pitch in Hz, loudness or decay in dB or seconds, reverberation time, stereo position, audio-visual onset alignment, silence in vacuum --- you must call the relevant tool before producing the verdict for that statement. For statements that are purely qualitative (e.g.\ timbre matching a real-world source class), tool use is at your discretion.

You may call multiple tools across multiple turns. Pass the path \texttt{\{video\_path\}} to all tool calls.

\textbf{Required minimum tool coverage for this clip}: before you produce any verdict, you must have called \textbf{at least one audio tool} (\texttt{dsp\_*}). The call should target a measurable acoustic quantity informative for the physical commonsense statements being judged. Do not call tools whose output you will not actually use.
\end{quote}

\subsection{Stage~2: verdict prompt}
\label{app:prompts_verdict}

The verdict prompt instantiates the five-dimension rubric of Section~\ref{subsec:rubric} for the current prompt. The Semantic-Adherence (SA) block hard-codes the visible objects/event and the audible objects/sound from the rubric's \texttt{basic\_standards}; the Physical-Commonsense (PC) block enumerates the per-prompt yes/no statements written into \texttt{key\_standards.video\_pc}, \texttt{key\_standards.audio\_pc}, and \texttt{key\_standards.av\_pc}. A boolean \texttt{flags.silence\_expected} switches the \asa\ wording to a silence check rather than an audibility check, so silence-by-design prompts are not penalised for not being audible.

\begin{quote}\small
Suppose you are an expert in summarization and finding answers. Here is the text description from another large language model about an AI-generated audio-video clip:

``\{stage-1 description\}''

Based on this description, please answer each of the following questions with strictly ``Yes'' or ``No''.

\textbf{Basic Standards (Semantic Adherence)}
\begin{enumerate}[leftmargin=1.5em, itemsep=0pt, label=\arabic*.]
\item \textbf{video\_sa.objects} --- Are all of the following visually present in the clip: \{video.objects\}? Answer Yes or No.
\item \textbf{video\_sa.event} --- Is the event ``\{video.event\}'' visually depicted in the clip? Answer Yes or No.
\item \textbf{audio\_sa.objects} --- Are the sound source(s) \{audio.objects\} audible in the clip? \emph{(when} \texttt{silence\_expected}\emph{: ``would normally be audible if real-world physics held; answer Yes if they are appropriately represented as such (typically silent here)'')}
\item \textbf{audio\_sa.sound} --- Is the sound \{audio.sound\} clearly audible in the clip? \emph{(when} \texttt{silence\_expected}\emph{: ``the clip is expected to be silent during the depicted event; answer Yes if it is appropriately silent throughout with no audible leak-through'')}
\end{enumerate}

\textbf{Key Standards (Physical Commonsense)}

Check whether each of the following physics statements is true of the clip. Answer ``Yes'' if the statement is clearly true; ``No'' if it is false, ambiguous, or only partially true.
\begin{itemize}[leftmargin=1.5em, itemsep=0pt, label={--}]
\item \textbf{video\_pc.Statement\_$i$}: \{rubric \texttt{key\_standards.video\_pc[i]}\}
\item \textbf{audio\_pc.Statement\_$i$}: \{rubric \texttt{key\_standards.audio\_pc[i]}\}
\item \textbf{av\_pc.Statement\_$i$}: \{rubric \texttt{key\_standards.av\_pc[i]}\}
\end{itemize}

\textbf{Output}

Return JSON with one entry in \texttt{per\_statement} for every statement id listed above. Each entry has \texttt{statement\_id}, \texttt{observation} (1--3 sentences citing the description), and \texttt{verdict} (``Yes'' or ``No'').
\end{quote}

The call uses \texttt{response\_mime\_type = "application/json"} and a Pydantic-derived JSON schema that pins \texttt{verdict} to the literal set \{\texttt{"Yes"}, \texttt{"No"}\}. If the returned JSON is missing any of the expected statement ids, the agent retries once with an explicit ``\texttt{per\_statement} must contain exactly one entry for EACH of these ids: \dots'' addendum; if parsing or coverage still fails, the verdict for the missing statements defaults to \texttt{No} and the run is flagged with a \texttt{parse\_error} field in the released JSON. Per-aspect aggregation is then strict-AND across statements (Section~\ref{subsec:rubric}).

\subsection{MLLM-as-judge baseline prompt}
\label{app:prompts_mllm}

The \emph{MLLM-as-judge baseline} row in Table~\ref{tab:human_consistency} uses the verdict-stage SA and PC blocks above, but with no Stage~1 observation and no tool block; the model is asked to watch and listen to the clip and answer each statement directly:

\begin{quote}\small
Watch and listen to the clip. For each statement below, return verdict ``Yes'' or ``No''.

\{SA block + PC block as above\}

Return JSON with one entry in \texttt{per\_statement} for each statement id; each entry has \texttt{statement\_id} and \texttt{verdict}.
\end{quote}

The baseline therefore differs from \agent in exactly two places: it has no ReAct stage, and the verdict schema omits the per-statement \texttt{observation} field. All other settings (backbone, decoding configuration, MP4 inline upload, JSON schema enforcement, retry-on-incomplete-coverage, strict-AND aggregation) are identical, isolating the contribution of the ReAct loop and the DSP tools.

\section{\agent Agreement and Correlation with Human Ratings}
\label{app:agent_correlation}

This appendix complements the agent-evaluation results in Section~\ref{subsec:agent} by reporting the ranking fidelity of every automatic evaluator against the human-majority labels collected on the 268 physics-following prompts of \bench. Ranking fidelity asks whether the evaluator preserves the same per-(model, dimension) pass rates that humans produce, and is the appropriate target for an evaluator whose primary use is to rank joint audio-video generators. Cell-wise binary agreement is reported per-dimension in Appendix~\ref{app:agent_tool_ablation}.

Table~\ref{tab:agent_correlation_l1} reports the Pearson correlation between every evaluator and human-majority labels on the 35 (model, dimension) pass rates that drive the leaderboard in Table~\ref{tab:human_eval_main}. The headline \agent achieves $r = 0.934$, compared to $r = 0.890$ for the tool-free MLLM-as-judge baseline. Per-dimension correlations show that the gain is concentrated on the audio and cross-modal dimensions, where the deterministic acoustic measurements add evidence beyond the multimodal language model's native perception: \asa $0.883 \to 0.988$, \apc $0.909 \to 0.967$, \avpc $0.965 \to 0.966$. Visual dimensions, on which the underlying multimodal language model already perceives strongly, remain at the same level (\vsa $0.990 \to 0.985$, \vpc $0.983 \to 0.971$).

\begin{table}[h]
\centering
\small
\setlength{\tabcolsep}{6pt}
\renewcommand{\arraystretch}{1.1}
\begin{tabular}{l c ccccc}
\toprule
\textbf{Method} & \textbf{Overall (n=35)} & \vsa & \asa & \vpc & \apc & \avpc \\
\midrule
MLLM-as-judge baseline             & 0.890          & 0.990          & 0.883          & 0.983          & 0.909          & 0.965 \\
Agent with audio tools (\agent)    & \textbf{0.934} & 0.985          & \textbf{0.988} & 0.971          & \textbf{0.967} & \textbf{0.966} \\
Agent with visual tools            & 0.888          & 0.978          & 0.962          & 0.938          & 0.923          & 0.912 \\
Agent with audio + visual tools    & 0.847          & 0.956          & 0.928          & 0.956          & 0.944          & 0.940 \\
\bottomrule
\end{tabular}
\caption{\textbf{Ranking fidelity: Pearson $r$ between automatic evaluators and human-majority labels on (model, dimension) pass rates.} Overall is computed across the full 35 (model, dimension) cells; per-dimension columns each correlate seven (one per generator) (evaluator, human) pass-rate pairs.}
\label{tab:agent_correlation_l1}
\end{table}

All four variants achieve $r \geq 0.847$ at the overall granularity, indicating that every evaluator preserves the human-derived generator ordering. Among them, the audio-tool \agent is the only configuration that improves over the MLLM-as-judge baseline. We discuss why the visual-tool and audio-plus-visual-tool variants fail to improve in Appendix~\ref{app:agent_tool_ablation}.

\paragraph{Anti-AV-Physics prompts.} We also compare \agent with the human panel on the 53 Anti-AV-Physics prompts for the three proprietary models (Table~\ref{tab:agent_anti}). \agent agrees with the human-majority label on 0.717 of the (clip, dimension) cells, and their pass rates correlate at $r=0.832$ over the 15 (model, dimension) pairs. Both find that few clips generate the requested violation.

\begin{table}[h]
\centering
\small
\setlength{\tabcolsep}{8pt}
\begin{tabular}{lcc}
\toprule
\textbf{Model} & \textbf{Human PC} & \textbf{\agent PC} \\
\midrule
Seedance 2.0   & 0.208 & 0.132 \\
Kling 3.0 Omni & 0.189 & 0.038 \\
Veo 3.1        & 0.113 & 0.075 \\
\bottomrule
\end{tabular}
\caption{\textbf{Anti-AV-Physics PC from the human panel and \agent,} over the 53 Anti-AV-Physics prompts.}
\label{tab:agent_anti}
\end{table}

\section{\agent Tool Ablation}
\label{app:agent_tool_ablation}

Table~\ref{tab:agent_tool_ablation} extends the main-text agreement comparison (Table~\ref{tab:human_consistency}) to four evaluator configurations that share the same Gemini~3.1 Pro Preview backbone and the same ReAct loop, but differ only in their tool inventory: (i) the \emph{MLLM-as-judge baseline} with no tools; (ii) Agent with audio DSP tools (LUFS, RT60, F0, onset, etc.), which is the headline \agent; (iii) Agent with visual tools (frame extraction, cropping, counting); and (iv) Agent with the union of audio and visual tools. Agreement is reported per (model, prompt, dimension) cell after strict-AND aggregation, on the same 9{,}380 cells covered by all four evaluators.

\begin{table}[h]
\centering
\small
\setlength{\tabcolsep}{6pt}
\renewcommand{\arraystretch}{1.1}
\begin{tabular}{l cccccc}
\toprule
\textbf{Method} & \vsa & \asa & \vpc & \apc & \avpc & \textbf{Avg. $\pm$ std} \\
\midrule
MLLM-as-judge baseline                  & 0.803          & 0.724          & 0.753          & 0.616          & 0.685          & 0.716 $\pm$ 0.071 \\
Agent with audio tools (\agent)         & \textbf{0.830} & \textbf{0.770} & \textbf{0.797} & \textbf{0.768} & \textbf{0.768} & \textbf{0.787 $\pm$ 0.027} \\
Agent with visual tools                 & 0.789          & 0.713          & 0.752          & 0.716          & 0.748          & 0.744 $\pm$ 0.031 \\
Agent with audio + visual tools         & 0.761          & 0.705          & 0.738          & 0.739          & 0.756          & 0.740 $\pm$ 0.022 \\
\bottomrule
\end{tabular}
\caption{\textbf{\agent tool ablation: per-dimension agreement with human-majority labels.} Agreement is measured per (model, prompt, dimension) cell after strict-AND aggregation, identical to Table~\ref{tab:human_consistency}. Avg.\ $\pm$ std is the sample mean and standard deviation across the five rubric dimensions.}
\label{tab:agent_tool_ablation}
\end{table}

\paragraph{Why audio tools help.} The MLLM-as-judge baseline is weakest on the two physics-sensitive dimensions whose answers depend on quantitative acoustic evidence: \apc at 0.616 and \avpc at 0.685. These are exactly the dimensions where a free-form multimodal perception cannot reliably estimate sub-second timing or fractional-dB loudness. The audio DSP toolchain replaces that estimation with a deterministic measurement: LUFS for amplification, RT60 for enclosure, onset alignment for cross-modal causality, F0 for pitch, and stereo balance for source lateralization. The largest gains land where they should, with \apc up by $0.152$ and \avpc by $0.083$. The dimension-wise standard deviation also collapses from $0.071$ to $0.027$, which means the gains lift the weak dimensions without disturbing the visual ones.

\paragraph{Why visual tools help less.} The underlying multimodal language model already sees the video clearly: it scores $0.803$ on \vsa and $0.753$ on \vpc with no tools at all. Frame extraction, cropping, and counting therefore mostly resurface evidence the model has already encoded natively, and they pay two costs for it. First, every tool call consumes a ReAct turn that could have gone toward the verdict. Second, a cropped or single-frame view occasionally disagrees with the model's whole-clip impression and pulls the verdict toward the local view. The result is that the visual-tool variant sits strictly below \agent on every dimension, including \vsa from $0.830$ to $0.789$ and \vpc from $0.797$ to $0.752$.

\paragraph{Why audio + visual tools help less.} The combined toolchain keeps the audio measurements but adds the visual-tool overhead on every cell. Each visual call still spends a ReAct turn, still risks disagreeing with the model's native vision, and now competes with the audio measurements for the model's attention budget on the way to the verdict. The audio-side gains are consequently diluted: \apc drops from $0.768$ to $0.739$, \avpc from $0.768$ to $0.756$, and \asa from $0.770$ to $0.705$. The combined configuration finishes lower than \agent on every dimension. We therefore use the audio-tool configuration as the headline \agent in the main text. The complementary cell-wise metrics on the same cells also place the audio-tool \agent first: cell-wise Pearson $r$ of $0.560$ and agreement of $0.787$, against $0.471 / 0.716$ for the MLLM-as-judge baseline, $0.470 / 0.744$ for the visual-tool variant, and $0.462 / 0.740$ for the audio + visual variant.

\paragraph{Significance.} Table~\ref{tab:agent_paired} compares \agent with the other three methods on the same cells. The ranking-fidelity values of Table~\ref{tab:agent_correlation_l1} lead to the same conclusion: \agent leads the baseline by 0.044 in Pearson $r$, the agent with visual tools by 0.046, and the agent with audio and visual tools by 0.086, and none of 2,000 prompt resamples reverses these orderings. The 0.002 between the baseline and the agent with visual tools has a 95\% CI of $[-0.029, +0.033]$, so these two are not distinguishable in $r$.

\begin{table}[h]
\centering
\small
\begin{tabular}{@{}lccc@{}}
\toprule
\textbf{\agent vs.} & \textbf{Difference} & \textbf{95\% CI} & \textbf{McNemar $p$} \\
\midrule
MLLM-as-judge baseline          & $+0.071$ & $[+0.056, +0.088]$ & $5.3 \times 10^{-39}$ \\
Agent with visual tools         & $+0.043$ & $[+0.033, +0.055]$ & $1.0 \times 10^{-21}$ \\
Agent with audio + visual tools & $+0.047$ & $[+0.037, +0.057]$ & $3.5 \times 10^{-26}$ \\
\bottomrule
\end{tabular}
\caption{\textbf{Paired comparison of \agent with the other three methods.} Difference is the agreement of \agent minus the agreement of the other method over 9,380 cells. The interval comes from 2,000 prompt resamples, and $p$ is from a two-sided exact McNemar test.}
\label{tab:agent_paired}
\end{table}

\paragraph{Consistency across runs.} We repeat the evaluation a second time under the same configuration (Table~\ref{tab:agent_runs}). Between the two runs, \agent changes 11.9\% of its individual judgments, but these changes do not push the scores in one direction: no standard deviation exceeds 0.011, no entry moves by more than 0.016, and \agent stays first on every dimension in both runs.

\begin{table}[h]
\centering
\small
\resizebox{\textwidth}{!}{%
\begin{tabular}{l cccccc}
\toprule
\textbf{Method} & \vsa & \asa & \vpc & \apc & \avpc & \textbf{Avg.} \\
\midrule
MLLM-as-judge baseline          & 0.800 $\pm$ 0.004 & 0.721 $\pm$ 0.003 & 0.755 $\pm$ 0.003 & 0.618 $\pm$ 0.003 & 0.682 $\pm$ 0.005 & 0.715 $\pm$ 0.001 \\
Agent with audio tools (\agent) & \textbf{0.823 $\pm$ 0.011} & \textbf{0.770 $\pm$ 0.001} & \textbf{0.795 $\pm$ 0.003} & \textbf{0.767 $\pm$ 0.001} & \textbf{0.771 $\pm$ 0.005} & \textbf{0.785 $\pm$ 0.002} \\
Agent with visual tools         & 0.781 $\pm$ 0.011 & 0.712 $\pm$ 0.000 & 0.748 $\pm$ 0.006 & 0.716 $\pm$ 0.001 & 0.750 $\pm$ 0.003 & 0.741 $\pm$ 0.003 \\
Agent with audio + visual tools & 0.762 $\pm$ 0.002 & 0.707 $\pm$ 0.003 & 0.739 $\pm$ 0.001 & 0.742 $\pm$ 0.004 & 0.756 $\pm$ 0.000 & 0.741 $\pm$ 0.002 \\
\bottomrule
\end{tabular}}
\caption{\textbf{Agreement with human-majority labels over two runs of the same configuration,} as mean $\pm$ standard deviation across the runs, over the 268 physics-following prompts.}
\label{tab:agent_runs}
\end{table}

\paragraph{Open-source backbone.} We repeat all four configurations with Qwen3-Omni-30B-A3B-Instruct~\cite{xu2025qwen3omni}, served locally by vLLM at temperature 0. Prompts, tools, and aggregation are unchanged, and the audio is passed as a separate 16\,kHz mono track. Table~\ref{tab:qwen} reports the first of two runs. In both runs, \agent is first on every dimension and the baseline has the lowest average.

\begin{table}[h]
\centering
\small
\setlength{\tabcolsep}{8pt}
\begin{tabular}{l cccccc}
\toprule
\textbf{Method} & \vsa & \asa & \vpc & \apc & \avpc & \textbf{Avg.} \\
\midrule
MLLM-as-judge baseline          & 0.644 & 0.589 & 0.447 & 0.331 & 0.338 & 0.470 \\
Agent with audio tools (\agent) & \textbf{0.788} & \textbf{0.688} & \textbf{0.747} & \textbf{0.673} & \textbf{0.689} & \textbf{0.717} \\
Agent with visual tools         & 0.736 & 0.664 & 0.664 & 0.566 & 0.586 & 0.643 \\
Agent with audio + visual tools & 0.742 & 0.644 & 0.655 & 0.534 & 0.588 & 0.633 \\
\bottomrule
\end{tabular}
\caption{\textbf{Agreement with human-majority labels on the Qwen3-Omni backbone,} over the 268 physics-following prompts.}
\label{tab:qwen}
\end{table}

\section{Dataset Statistics}
\label{app:stats}

\bench contains 321 hand-authored prompts organized along the scene-evolution taxonomy of Section~\ref{subsec:taxonomy}: 268 physics-following prompts that exercise audio-visual physics under a specific kind of scene dynamics, plus 53 Anti-AV-Physics control prompts that ask the model to render a deliberate violation. Table~\ref{tab:dataset_stats} reports per-subcategory counts under the three scene categories: Steady State (C1), Event Transition (C2), and Environment Transition (C3). Each scene category contains three physics-following subcategories and a fourth Anti-AV-Physics subcategory.

\begin{table}[h]
\centering
\small
\setlength{\tabcolsep}{6pt}
\renewcommand{\arraystretch}{1.1}
\begin{tabular}{@{}ll l c@{}}
\toprule
\textbf{Scene Category} & \textbf{Code} & \textbf{Subcategory} & \textbf{N} \\
\midrule
\multirow{4}{*}{\shortstack[l]{C1: Steady State\\(159 prompts)}}
 & C1.1 & Source Material      & 54 \\
 & C1.2 & Source Anchoring     & 26 \\
 & C1.3 & Sound Persistence    & 47 \\
 & C1.4 & Anti-AV-Physics      & 32 \\
\midrule
\multirow{4}{*}{\shortstack[l]{C2: Event Transition\\(119 prompts)}}
 & C2.1 & Source Body          & 48 \\
 & C2.2 & Source Excitation    & 24 \\
 & C2.3 & Source Radiation     & 33 \\
 & C2.4 & Anti-AV-Physics      & 14 \\
\midrule
\multirow{4}{*}{\shortstack[l]{C3: Environment Transition\\(43 prompts)}}
 & C3.1 & Propagation Medium   & 13 \\
 & C3.2 & Enclosure Geometry   &  9 \\
 & C3.3 & Sound Attenuation    & 14 \\
 & C3.4 & Anti-AV-Physics      &  7 \\
\midrule
\multicolumn{3}{l}{\textbf{Total physics-following}} & \textbf{268} \\
\multicolumn{3}{l}{\textbf{Total Anti-AV-Physics}}   & \textbf{53}  \\
\multicolumn{3}{l}{\textbf{Total prompts}}           & \textbf{321} \\
\bottomrule
\end{tabular}
\caption{Per-scene-category and per-subcategory prompt counts in \bench. The fourth subcategory of every scene category, marked \emph{Anti-AV-Physics}, holds out the corresponding control set.}
\label{tab:dataset_stats}
\end{table}

Each prompt is paired with a per-prompt rubric instance instantiated from the five-dimension rubric of Section~\ref{subsec:rubric}. Across the 321 prompts the rubric set contains 2{,}763 individual Y/N statements, averaging 8.6 statements per prompt (range 7--13), with per-dimension averages of 2.00 \vsa, 2.00 \asa, 1.50 \vpc, 1.49 \apc, and 1.61 \avpc statements per prompt. Each of the seven generators is evaluated on every prompt, and each generated clip is rated by three independent annotators, yielding $321 \times 7 \times 3 \approx 6{,}700$ rubric-instance ratings and $\approx 58{,}000$ statement-level human verdicts in total. Annotator pool composition and inter-rater agreement are detailed in Appendix~\ref{app:human_protocol}.

\section{Design Principles}
\label{app:principles}

We describe the principles that guide \bench's design.

\textbf{Measurability.} Every taxonomy subcategory maps to at least one deterministic acoustic measurement that can serve as objective evidence (Appendix Table~\ref{tab:full_taxonomy}).

\textbf{Completeness.} The three scene categories span both static and dynamic facets of within-clip acoustic physics, with the Anti-AV-Physics control probing intentional violation.

\textbf{Non-redundancy.} Subcategories within each scene category test distinct physical principles, and a deterministic mapping between the primary and secondary taxonomies ensures that no prompt is double-counted.

\end{document}